\documentclass[prd,reprint,superscriptaddress,showpacs,amssymb,amsmath,amsfonts,aps]{revtex4-1} 

\usepackage{amsmath}
\usepackage{graphicx}
\usepackage{subfigure}
\usepackage{hyperref}
\usepackage{dcolumn}
\usepackage{bm}
\usepackage[retainorgcmds]{IEEEtrantools}
\usepackage{float}

\begin{document}

\title{Beam-spin Asymmetries from Semi-inclusive Pion Electroproduction}

\newcommand*{\ANL}{Argonne National Laboratory, Argonne, Illinois 60439}
\newcommand*{\ANLindex}{1}
\affiliation{\ANL}
\newcommand*{\ASU}{Arizona State University, Tempe, Arizona 85287-1504}
\newcommand*{\ASUindex}{2}
\affiliation{\ASU}
\newcommand*{\CSUDH}{California State University, Dominguez Hills, Carson, CA 90747}
\newcommand*{\CSUDHindex}{3}
\affiliation{\CSUDH}
\newcommand*{\CANISIUS}{Canisius College, Buffalo, NY 14208}
\newcommand*{\CANISIUSindex}{4}
\affiliation{\CANISIUS}
\newcommand*{\CMU}{Carnegie Mellon University, Pittsburgh, Pennsylvania 15213}
\newcommand*{\CMUindex}{5}
\affiliation{\CMU}
\newcommand*{\CUA}{Catholic University of America, Washington, D.C. 20064}
\newcommand*{\CUAindex}{6}
\affiliation{\CUA}
\newcommand*{\SACLAY}{CEA, Centre de Saclay, Irfu/Service de Physique Nucl\'eaire, 91191 Gif-sur-Yvette, France}
\newcommand*{\SACLAYindex}{7}
\affiliation{\SACLAY}
\newcommand*{\CNU}{Christopher Newport University, Newport News, Virginia 23606}
\newcommand*{\CNUindex}{8}
\affiliation{\CNU}
\newcommand*{\UCONN}{University of Connecticut, Storrs, Connecticut 06269}
\newcommand*{\UCONNindex}{9}
\affiliation{\UCONN}
\newcommand*{\EDINBURGH}{Edinburgh University, Edinburgh EH9 3JZ, United Kingdom}
\newcommand*{\EDINBURGHindex}{10}
\affiliation{\EDINBURGH}
\newcommand*{\FU}{Fairfield University, Fairfield CT 06824}
\newcommand*{\FUindex}{11}
\affiliation{\FU}
\newcommand*{\FIU}{Florida International University, Miami, Florida 33199}
\newcommand*{\FIUindex}{12}
\affiliation{\FIU}
\newcommand*{\FSU}{Florida State University, Tallahassee, Florida 32306}
\newcommand*{\FSUindex}{13}
\affiliation{\FSU}
\newcommand*{\GWUI}{The George Washington University, Washington, DC 20052}
\newcommand*{\GWUIindex}{14}
\affiliation{\GWUI}
\newcommand*{\ISU}{Idaho State University, Pocatello, Idaho 83209}
\newcommand*{\ISUindex}{15}
\affiliation{\ISU}
\newcommand*{\INFNFE}{INFN, Sezione di Ferrara, 44100 Ferrara, Italy}
\newcommand*{\INFNFEindex}{16}
\affiliation{\INFNFE}
\newcommand*{\INFNFR}{INFN, Laboratori Nazionali di Frascati, 00044 Frascati, Italy}
\newcommand*{\INFNFRindex}{17}
\affiliation{\INFNFR}
\newcommand*{\INFNGE}{INFN, Sezione di Genova, 16146 Genova, Italy}
\newcommand*{\INFNGEindex}{18}
\affiliation{\INFNGE}
\newcommand*{\INFNRO}{INFN, Sezione di Roma Tor Vergata, 00133 Rome, Italy}
\newcommand*{\INFNROindex}{19}
\affiliation{\INFNRO}
\newcommand*{\ORSAY}{Institut de Physique Nucl\'eaire ORSAY, Orsay, France}
\newcommand*{\ORSAYindex}{20}
\affiliation{\ORSAY}
\newcommand*{\ITEP}{Institute of Theoretical and Experimental Physics, Moscow, 117259, Russia}
\newcommand*{\ITEPindex}{21}
\affiliation{\ITEP}
\newcommand*{\JMU}{James Madison University, Harrisonburg, Virginia 22807}
\newcommand*{\JMUindex}{22}
\affiliation{\JMU}
\newcommand*{\KNU}{Kyungpook National University, Daegu 702-701, Republic of Korea}
\newcommand*{\KNUindex}{23}
\affiliation{\KNU}
\newcommand*{\LPSC}{LPSC, Universite Joseph Fourier, CNRS/IN2P3, INPG, Grenoble, France
}
\newcommand*{\LPSCindex}{24}
\affiliation{\LPSC}
\newcommand*{\UNH}{University of New Hampshire, Durham, New Hampshire 03824-3568}
\newcommand*{\UNHindex}{25}
\affiliation{\UNH}
\newcommand*{\NSU}{Norfolk State University, Norfolk, Virginia 23504}
\newcommand*{\NSUindex}{26}
\affiliation{\NSU}
\newcommand*{\OHIOU}{Ohio University, Athens, Ohio  45701}
\newcommand*{\OHIOUindex}{27}
\affiliation{\OHIOU}
\newcommand*{\ODU}{Old Dominion University, Norfolk, Virginia 23529}
\newcommand*{\ODUindex}{28}
\affiliation{\ODU}
\newcommand*{\RPI}{Rensselaer Polytechnic Institute, Troy, New York 12180-3590}
\newcommand*{\RPIindex}{29}
\affiliation{\RPI}
\newcommand*{\URICH}{University of Richmond, Richmond, Virginia 23173}
\newcommand*{\URICHindex}{30}
\affiliation{\URICH}
\newcommand*{\ROMAII}{Universita' di Roma Tor Vergata, 00133 Rome Italy}
\newcommand*{\ROMAIIindex}{31}
\affiliation{\ROMAII}
\newcommand*{\MSU}{Skobeltsyn Institute of Nuclear Physics, Lomonosov Moscow State University, 119234 Moscow, Russia}
\newcommand*{\MSUindex}{32}
\affiliation{\MSU}
\newcommand*{\SCAROLINA}{University of South Carolina, Columbia, South Carolina 29208}
\newcommand*{\SCAROLINAindex}{33}
\affiliation{\SCAROLINA}
\newcommand*{\JLAB}{Thomas Jefferson National Accelerator Facility, Newport News, Virginia 23606}
\newcommand*{\JLABindex}{34}
\affiliation{\JLAB}
\newcommand*{\UTFSM}{Universidad T\'{e}cnica Federico Santa Mar\'{i}a, Casilla 110-V Valpara\'{i}so, Chile}
\newcommand*{\UTFSMindex}{35}
\affiliation{\UTFSM}
\newcommand*{\GLASGOW}{University of Glasgow, Glasgow G12 8QQ, United Kingdom}
\newcommand*{\GLASGOWindex}{36}
\affiliation{\GLASGOW}
\newcommand*{\VIRGINIA}{University of Virginia, Charlottesville, Virginia 22901}
\newcommand*{\VIRGINIAindex}{37}
\affiliation{\VIRGINIA}
\newcommand*{\WM}{College of William and Mary, Williamsburg, Virginia 23187-8795}
\newcommand*{\WMindex}{38}
\affiliation{\WM}
\newcommand*{\YEREVAN}{Yerevan Physics Institute, 375036 Yerevan, Armenia}
\newcommand*{\YEREVANindex}{39}
\affiliation{\YEREVAN}

\newcommand*{\NOWORSAY}{Institut de Physique Nucl\'eaire ORSAY, Orsay, France}
\newcommand*{\NOWUTFSM}{Universidad T\'{e}cnica Federico Santa Mar\'{i}a, Casilla 110-V Valpara\'{i}so, Chile}
\newcommand*{\NOWROMAII}{Universita' di Roma Tor Vergata, 00133 Rome Italy}
\newcommand*{\NOWUK}{University of Kentucky, Lexington, KY 40506}
\newcommand*{\NOWMATH}{Mathworks, Natick, MA 01760}
\newcommand*{\NOWODU}{Old Dominion University, Norfolk, VA 23529}


\author{W.~Gohn}
\email{gohn@pa.uky.edu}
\altaffiliation[Current address:]{\NOWUK}
\affiliation{\UCONN}


\author{H.~Avakian}
\affiliation{\JLAB}


\author{K.~Joo}
\affiliation{\UCONN}


\author{M.~Ungaro}
\affiliation{\JLAB}

\author{K.P.~Adhikari}
\affiliation{\ODU}
\author {M.~Aghasyan} 
\affiliation{\INFNFR}
\author {M.J.~Amaryan} 
\affiliation{\ODU}
\author {M.D.~Anderson} 
\affiliation{\GLASGOW}
\author {S. ~Anefalos~Pereira} 
\affiliation{\INFNFR}
\author {J.~Ball} 
\affiliation{\SACLAY}
% ...

\collaboration{The CLAS Collaboration}
\noaffiliation

\begin{abstract}
We have measured the moment $A_{LU}^{\sin\phi}$ corresponding to the polarized electron beam-spin asymmetry in SIDIS. $A_{LU}^{\sin\phi}$ is a twist-3 quantity providing information about quark-gluon correlations. Data were taken with the CLAS Spectrometer at Jefferson Lab using a 5.498 GeV longitudinally polarized electron beam and an unpolarized liquid hydrogen target. All three pion channels ($\pi^+$, $\pi^0$ and $\pi^-$) were measured simultaneously over a large range of kinematics within the virtuality range $Q^2 \approx$ 1.0-4.5 GeV$^2$. The observable was measured with better than 1\% statistical precision over a large range of $z$, $P_T$, $x_B$, and $Q^{2}$, which permits comparison with several reaction models. The discussed measurements provide an upgrade in statistics over previous measurements, and serve as the first evidence for the negative sign of the $\pi^{-}$ $\sin\phi$ moment.
\end{abstract}

\keywords{SIDIS,beam-spin asymmetry,TMDs}

\maketitle

\section{Introduction}

For decades, experiments in deep-inelastic scattering (DIS) have mapped out the momentum distributions in the nucleon in terms of one-dimensional parton distribution functions (PDFs). While these measurements provided significant insight into the nucleon structure, some questions arose that could not be addressed in this one-dimensional picture. Most notably, the EMC experiment at CERN~\cite{emc} found that the quark-spin contribution to the proton's spin is only about 30$\%$. Recent measurements of the gluon contribution to the proton spin have shown this contribution to be too small to to saturate the spin sum rule~\cite{collaboration:2011fga,Adamczyk:2012qj,Adare:2008aa,deFlorian:2009vb}. These results necessitate an understanding of the orbital motion of quarks in the proton.

In recent years, the hadronic physics community has extended the investigation of partonic structure of hadrons beyond the collinear PDFs by exploring the parton's motion and its spatial distribution in the direction perpendicular to the parent hadron's momentum.  Two sets  of non-perturbative functions  have been introduced to investigate and 
describe the structure of hadrons at the quark-gluon level.
Transverse momentum dependent parton distributions (TMDs) carry information on the longitudinal and transverse momentum distributions. Generalized parton distributions (GPDs) carry information on the longitudinal momentum distribution and the transverse positions of the partons. GPDs and TMDs are connected through the Wigner distribution functions~\cite{Belitsky:2003nz} or Generalized Transverse Momentum Distributions (GTMDs)~\cite{Meissner:2008xs, Lorce:2011dv} and provide complete description of the three­dimensional structure of the nucleon.

Measurements of azimuthal moments, in particular
the single spin asymmetries (SSAs), have emerged recently as a powerful tool to probe the nucleon structure through measurements of GPDs and TMDs in hard exclusive and semi­inclusive electro­production of mesons and photon, respectively.
Pion electroproduction in semi-inclusive DIS (SIDIS), when a final state pion is detected with the final state lepton, is an important tool for studying the TMD of partons.
Assuming single photon exchange, the differential cross section for this process is written as a product of leptonic and hadronic tensors:

\begin{equation}
\frac{d\sigma}{dx_{B}dzdyd^{2}P_{T}}=\frac{\pi\alpha^{2}yz}{2Q^{4}}\mathcal{L_{\mu\nu}}2M\mathcal{W^{\mu\nu}},
\label{eq:dsig1}
\end{equation}

\noindent
where $x_B$ is the fraction of the proton's momentum that is carried by the struck quark,  $y$ is the energy fraction of the incoming lepton carried by the virtual photon, $z$ is the energy fraction of the virtual photon carried by the outgoing hadron, $P_{T}$ is the transverse momentum of the final state hadron, and $Q^{2}$ is the virtuality of the collision. The leptonic tensor is $\mathcal{L_{\mu\nu}}$ and the hadronic tensor is $2M\mathcal{W^{\mu\nu}}$. 

\begin{figure}
\centering
\includegraphics[width=6cm]{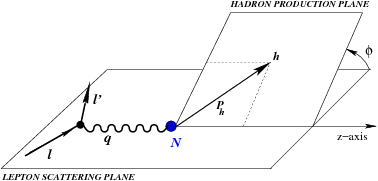}
\caption{The scattering diagram of the SIDIS process. The angle $\phi$ is the angle between the plane described by the incident and scattered electron and that defined by the direction of the emitted hadron.}
\label{fig:phi}
\end{figure}

The SIDIS cross section for single pion electroproduction may be expressed as a function of the angle $\phi$ between the leptonic and hadronic scattering planes, which is described by the Trento convention~\cite{Bacchetta:2004jz}, as shown in Fig.~\ref{fig:phi}. The expression for the cross section for an unpolarized target can be written as
\begin{equation}
\begin{split}
d\sigma = & d\sigma_{0}(1+A_{UU}^{\cos\phi}\cos\phi \\
& +A_{UU}^{\cos2\phi}\cos2\phi+\lambda_{e}A_{LU}^{\sin\phi}\sin\phi),
\end{split}
\label{eq:cs}
\end{equation}
 where $\lambda_{e} =\pm1$ is the electron beam helicity and $d\sigma_{0}$ is a normalization that will cancel from the asymmetry. The coefficients $A_{UU}^{\cos\phi}$, $A_{UU}^{\cos2\phi}$, and $A_{LU}^{\sin\phi}$ are the three moments of the cross section. The subscripts on the three coefficients denote the beam polarization first ($U$ represents unpolarized and $L$ represents longitudinally polarized) and target polarization second (all are $U$ for unpolarized). Each of the three moments in the cross section may be written in a model-independent way as a set of structure functions~\cite{Bacchetta:2006tn}, each of which is related to a specific set of TMDs and fragmentation functions.

The presented measurement is of the third moment of the above cross-section $A_{LU}^{\sin\phi}$, the only one of the three that is dependent on beam helicity. The helicity dependence of the $\sin\phi$ term arises from the antisymmetric part of the leptonic tensor and the way it couples to the hadronic tensor. This moment can be written as the ratio of two structure functions,
 where the denominator represents the beam spin independent part~\cite{Bacchetta:2006tn} and the numerator is the structure function $F_{LU}^{\sin\phi}$.

The  structure function $F_{LU}^{\sin\phi}$ is related to quark-gluon-quark correlations in the proton, and can be expressed as a convolutions of distribution and fragmentation functions~\cite{Levelt:1994np,Bacchetta:2006tn}. The twist of an operator is defined as the power with which the hadron mass scale $M$ occurs in a matrix element. Reference~\cite{Jaffe:1996zw} provides a working redefinition of twist to be two plus the power $p$ in which the ratio $M/Q$ occurs where $Q$ is the magnitude of the four momentum transferred (twist order =$2+p$). Twist-2 is called leading twist and twist-3 is subleading twist because these terms are suppressed by $O(M/Q)$. The twist-2 terms describe parton densities given by quark-antiquark correlators $\langle \bar q q \rangle$ and the twist-3 term is the correlation of quark and gluon fields described by quark-gluon-quark correlators $\langle \bar q G q \rangle$.  $F_{LU}^{\sin\phi}$ intrinsically has a  twist of order 3.

\begin{figure}
\includegraphics[width=6cm]{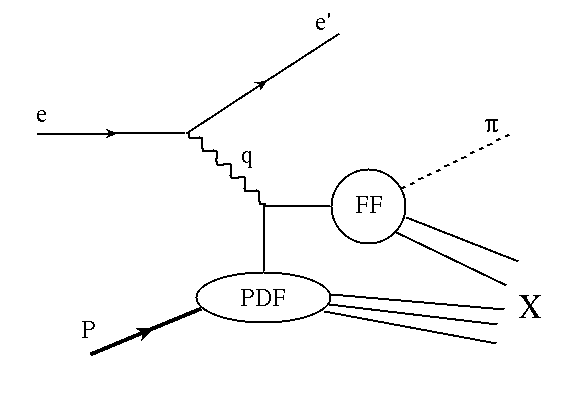}
\caption{SIDIS scattering process including the parton distribution function and fragmentation function.}
\label{fig:sidis_diagram}
\end{figure}

The structure function $F_{LU}^{\sin\phi}$ is composed of convolutions of parton distributions with fragmentation functions summed over quark flavor, as shown in Fig.~\ref{fig:sidis_diagram}:





\begin{equation}
\begin{split}
F_{LU}^{\sin\phi}  \propto & 
\frac{M}{Q}\,\sum\limits_ae_a^2\biggl( 
	  e^a \, H_1^{\perp a} 
   	+ f_1^a \tilde{G}^{\perp a} \\
   	&+ g^{\perp a} D_1^a 
   	+ h_1^{\perp a} \tilde{E}^a\biggr)
\end{split}
	\label{eq:flu}
\end{equation}

This structure function was the subject of numerous theoretical 
and phenomenological studies
\cite{Efremov:2002ut,Schweitzer:2003uy,Ohnishi:2003mf,Afanasev:2003ze,Yuan:2003gu,Bacchetta:2004zf,Metz:2004je,Pijlman:2006vm,Bacchetta:2006tn,Lu:2012gu,Mao:2012dk}.
Nevertheless, there is presently no satisfactory understanding
of how much each function in Eq.~\ref{eq:flu} contributes.
Of particular importance are the Boer-Mulders function $h_{1}^{\perp}$, which is a leading twist naive time-reversal odd TMD~\cite{Boer:1997nt}, and $g^{\perp}$, the twist-3 time reversal odd TMD~\cite{Bacchetta:2004zf}, which has been described as the higher-twist analog of the Sivers function~\cite{Sivers:1990fh}. $e(x)$ is a chiral-odd twist-3 PDF~\cite{Jaffe:1991ra}. It has been suggested that the $x^{2}$ moment of $e(x)$ could be related to the transverse force acting on the transversely polarized quarks in an unpolarized nucleon~\cite{Burkardt:2009rf}.

The $\sin\phi$ moment also provides access to the twist-3 fragmentation functions, $\tilde G^{\perp}$ and $\tilde E$, and the Collins function~\cite{Collins:1992kk}, $H_{1}^{\perp}$, which has previously been seen to cause asymmetries of opposite sign for oppositely charged pions at Belle~\cite{Seidl:2008xc}, HERMES~\cite{Airapetian:2010ds}, and COMPASS~\cite{Alekseev:2010rw}. 


Every term in the structure function can be shown to be a pure twist-3 term at leading order. Hence, the often used Wandzura-Wilczeck approximation, which neglects
all interaction dependent parts in twist-3 terms in a structure function, is not valid in this case as it would demand that the entire asymmetry to be zero, which is not the case.

A sizeable beam-spin asymmetry in semi-inclusive pion electroproduction was predicted by F. Yuan~\cite{Yuan:2003gu}, arising from the convolution of the  Boer-Mulders function with the twist-3 fragmentation function $\tilde E$~\cite{Jaffe:1991ra} or by the convolution of the twist-3 T-odd TMD $g^{\perp}$ with the fragmentation function $D_{1}$~\cite{Metz:2004je, Mao:2012dk}. An asymmetry arising from the convolution of the Collins function with the chiral odd twist-3 TMD $e(x)$ is described in~\cite{Schweitzer:2003uy}.

Recently, the twist-3 distributions have seen an increase in attention for their significant relationship to the quark orbital angular momentum~\cite{Penttinen:2000dg,Ji:2012ba,Hatta:2012cs}. Measurements of significant SSAs described by 
higher twist distribution and fragmentation functions indicate that multi-parton correlations are significant and their understanding is crucial for a complete description of the structure of the nucleon.
Measurements of $A_{LU}^{\sin\phi}$ should  provide access to twist-3 TMDs and fragmentation functions, improving our understanding of quark-gluon-quark correlators in the proton.

\section{Experiment}

The current measurement at Jefferson Lab utilized a 5.498 GeV polarized electron beam with an average beam polarization of 0.75$\pm$0.03, incident upon an unpolarized liquid hydrogen target. Semi-inclusive pion electroproduction $ep \rightarrow e\pi^{\pm,0}X$, where $X$ denotes undetected final-state hadrons, was observed, leading to asymmetry measurements with absolute uncertainties of 0.015 or less in all three pion channels ($\pi^+$, $\pi^0$ and $\pi^-$). Measuring in all three channels simultaneously is important to reduce the experiment's systematic uncertainty. The measurements were completed with the Continuous Electron Beam Accelerator Facility (CEBAF) Large Acceptance Spectrometer (CLAS) during the E1-f run period from April through June of 2003. An integrated luminosity of 21 $fb^{-1}$ was collected during the experiment.

The electron beam was provided by CEBAF, consisting of two linear accelerators that propel polarized electrons to a total energy of 5.498 GeV. The helicity of the electrons was flipped with a frequency of 33 Hz in order to minimize systematic effects. A half-wave plate was also used to periodically flip the helicity definitions between positive and negative in order to further negate any systematic uncertainty due to beam helicity. The beam polarization was measured frequently with a M$\o$ller polarimeter with a negligible statistical uncertainty.

\begin{figure}
\centering
\includegraphics[width=10cm]{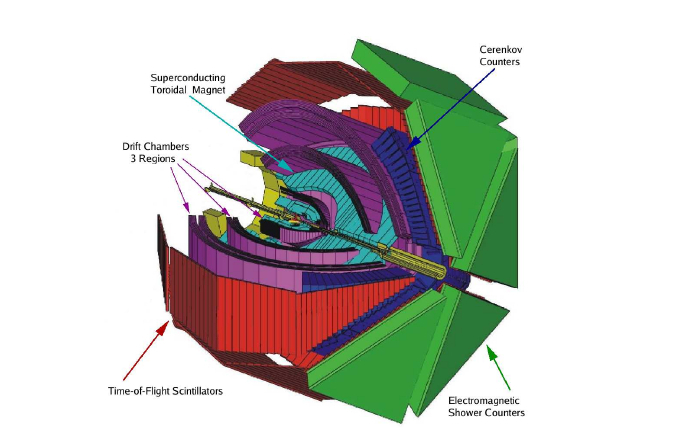}
\caption{(Color Online) The CLAS detector, including drift chambers, \v Cerenkov counters, electromagnetic calorimeters, and time-of-flight detectors.}
\label{fig:clas}
\end{figure}

The CLAS detector~\cite{clasnim}, shown in Fig.~\ref{fig:clas}, was located in Hall-B. The detector was composed of four detector sub-systems in a layered configuration and divided into six sectors in the azimuthal angle, providing nearly 4$\pi$ coverage. Particles were detected using drift chambers~\cite{dcnim} for tracking of charged particles, \v Cerenkov counters~\cite{ccnim} and electromagnetic calorimeters~\cite{ecnim} for electron identification, and time-of-flight detectors~\cite{tofnim} for hadron velocity measurements.

Charged particles traveled through CLAS in a curved trajectory due to a toroidal magnetic field. In the nominal configuration, many $\pi^{-}$ tracks were lost because they were bent out of the range of the CLAS acceptance. For E1-f, the CLAS torus magnet was run at 60$\%$ of its nominal current in order to maximize acceptance of $\pi^{-}$ tracks. 

\subsection{Electron identification}

Electrons were identified in the CLAS using a series of cuts on signals from the \v Cerenkov counter (CC) and electromagnetic calorimeter (EC). \v Cerenkov cuts were applied for tracks that did not produce a high number of photoelectrons in the CC. Timing and position information were used to discriminate between electrons and minimum ionizing tracks such as a $\pi^{-}$. These cuts were made in coincidence with cuts on the energy deposited in the EC, as well as position and timing information for each track. 

\begin{figure}
\centering
\includegraphics[width=8cm]{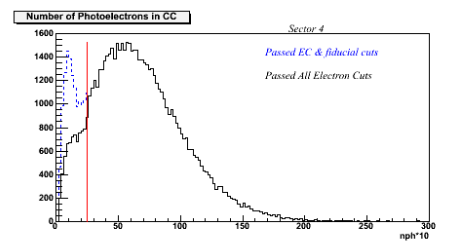}
\caption{(Color Online) Number of photoelectrons detected in the CLAS \v Cerenkov counter. Events with the number of photoelectrons $\times 10$ greater than 25 were kept, while those below were subjected to further cuts. The blue (dotted) histogram shows the total measured events passing cuts in the EC, but without additional CC cuts applied. The black (solid) histogram shows those events passing all electron identification cuts, including those imposed using the CC measurements when $N_{pe} \times 10 < 25$. The red line at $N_{pe} \times 10 = 25$ delineates the two regions with different CC cuts (see text).}
\label{fig:nph}
\end{figure}

The CLAS \v Cerenkov counter used photomultipliers to count photons emitted from \v Cerenkov radiation in the detector. A threshold of 2.5 photoelectrons was established, above which each track was accepted as an electron, as shown in Fig.~\ref{fig:nph}. If the number of photoelectrons ejected from the photocathode was less than 2.5, a series of other cuts were applied to the CC signal, including measurements of timing and position in the CC.

In the EC, a series of five cuts were used to discriminate between electrons and minimum ionizing particles, which were mostly negative pions. Cuts were made on the minimum momentum of an electron that could be seen due to the threshold of the calorimeter's trigger discriminator, which removed electron candidates with momentum below 0.6 GeV. Cuts were also made on the sampling fraction, the energy deposited in the pre-radiator (inner) and total absorption (outer) regions of the calorimeter, the location of each hit in the EC, and a matching of the timing between hits in the EC and time-of-flight detectors. 

\begin{figure}
\centering
\includegraphics[width=8cm]{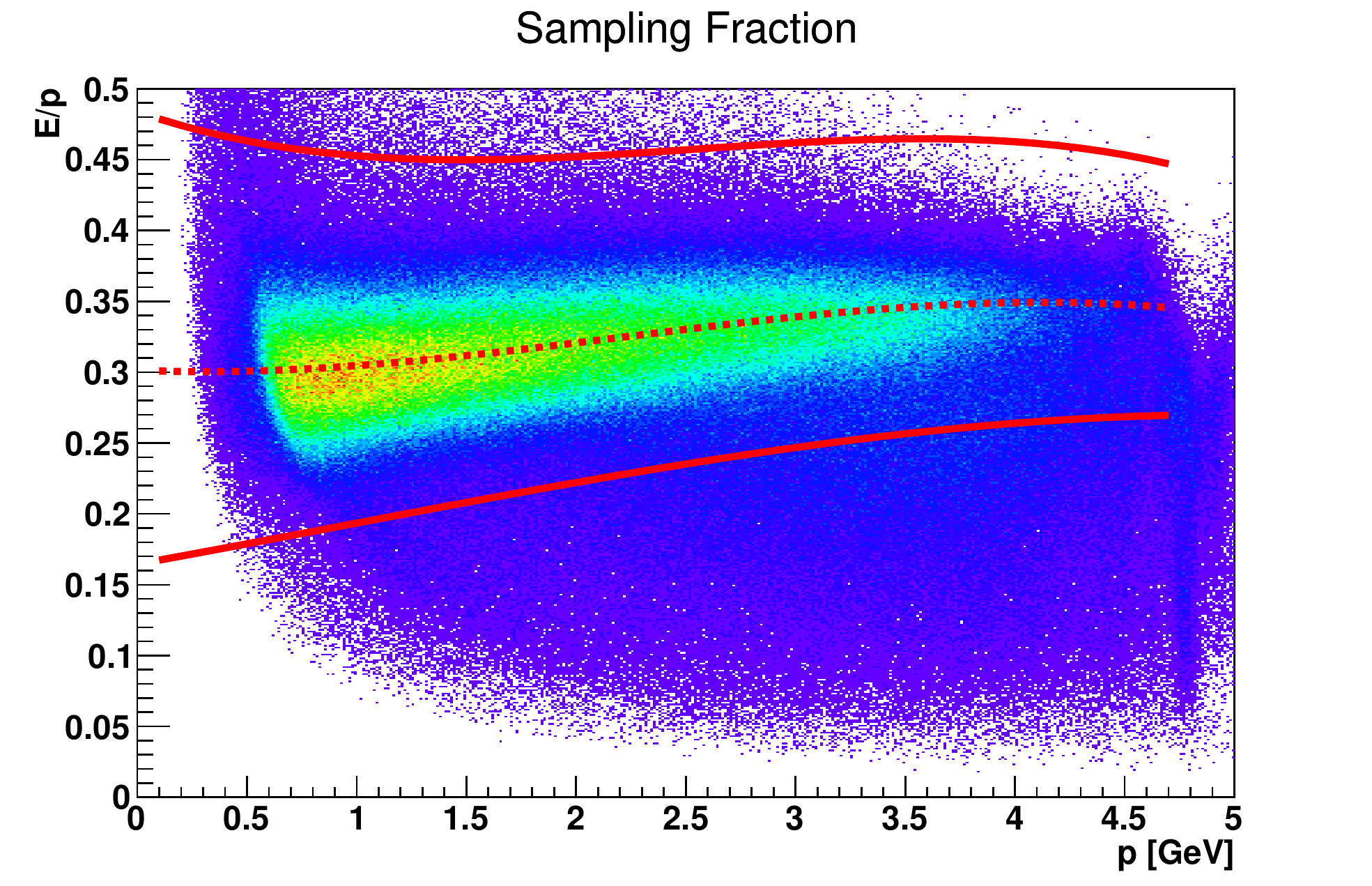}
\caption{(Color Online) Sampling fraction $E/p$ vs $p$ for electron candidates in the CLAS electromagnetic calorimeter. Only events between the two solid (red) lines were kept. The dotted line indicates the mean of the distribution.}
\label{fig:ecsf}
\end{figure}

Based on simple kinematic considerations, the sampling fraction, defined as the ratio of deposited energy over momentum as measured by the drift chambers, should be roughly constant in momentum. The sampling fraction was computed as a function of momentum in each of the six sectors of the CLAS, and each momentum slice was fit with a Gaussian to determine a mean and width of the distribution, as shown in Fig.~\ref{fig:ecsf}. The cut was then computed as a function of momentum as $\mu(p) + 3.5\sigma(p)/-3.0\sigma(p)$, where $\mu(p)$ is the mean of the Gaussian fits as a function of $p$ and $\sigma(p)$ is the function of the corresponding widths. The cut is asymmetric because there is a greater risk of pion contamination on the low side of the distribution than on the upper side.

\begin{figure}
\centering
\includegraphics[width=8cm]{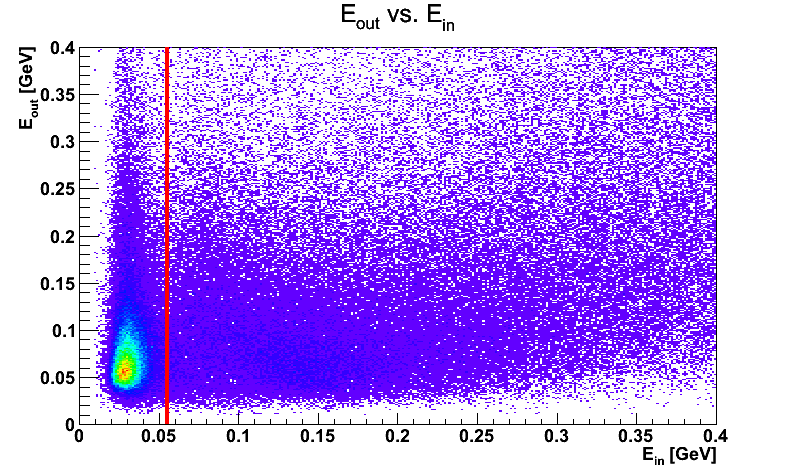}
\caption{(Color Online) A two-dimensional histogram of the energy deposited in the outer vs inner parts of the CLAS electromagnetic calorimeter. Electron candidates depositing less than 55 MeV (as represented by the red line) were removed.}
\label{fig:ein}
\end{figure}

The CLAS EC was separated into inner and outer regions. An electron passing through the EC would shower, depositing a large amount of energy into the EC. Pions on the other hand are minimum ionizing particles, which deposited a much smaller amount of energy that was easily separated by placing a cut on the total energy deposited in the inner part of the EC, keeping only events with $E_{\mbox{inner}}>55$ MeV, which is consistent with a 3$\sigma$ cut on the pion peak, as shown for one sector in Fig.~\ref{fig:ein}.

Geometric cuts were placed on the EC hits to remove events near the edges of each calorimeter, where the shower leakage may occur, resulting in an incomplete energy measurement. 

Cuts were made on the EC timing information by computing the difference in time between when the track hits the EC and when the same track hits the time-of-flight detector. Based on the known spatial separation between the detectors, this difference should be 0.7 ns. Events were accepted if their flight times were within 3$\sigma$ of a Gaussian fit to the $\Delta t$ distribution centered at 0.7 ns.

\subsection{Charged pion identification}

Charged hadrons were identified using the CLAS drift chambers and time-of-flight detectors. The drift chambers are used to measure the momentum of each track, and the particle's velocity was measured by the time-of-flight detector. These two measurements combined to give an accurate separation between the different charged particle tracks. The $\pi^{+}$ tracks had to be separated from protons and both charged pion channels were distinguished from kaons.

\begin{figure}
\centering
\includegraphics[width=8cm]{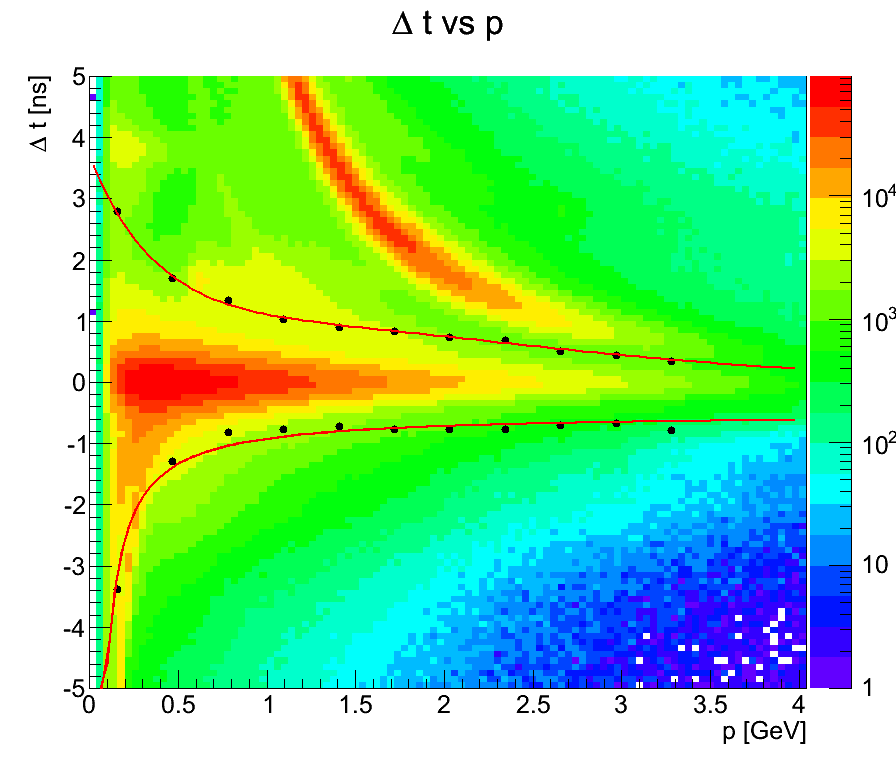}
\caption{(Color Online) $\Delta t$ vs momentum for positively charged tracks, where the cut shown removed proton and kaon tracks from the desired $\pi^{+}$ sample. The black dots define $\pm3\sigma$ for $\Delta t$ in each momentum bin, and the red curves are fits to those points, which were used to set maximum and minimum values on $\Delta t$. The plots for $\pi^{-}$ look similar but without the strong proton contribution.}
\label{fig:dt}
\end{figure}

To perform the required separation of charged pions from heavier charged tracks, a quantity $\Delta t$ was computed as the difference between the time recorded for each hit in the time-of-flight detector (corrected for the event start time) and the expected time for a pion to hit the detector as calculated from the track's momentum measured by the CLAS drift chambers. The calculated time is given by
\begin{equation}
t_{\mbox{calc}} = \frac{L_{\mbox{DC}}}{cp}\sqrt{p^{2}+m_{\pi}^2},
\end{equation}
where $L_{\mbox{DC}}$ is the path length of the track from the production vertex to the time­of­flight plane, $c$ is the speed of light, $p$ is the track's momentum, and $m_{\pi}$ is the known pion mass. Fig.~\ref{fig:dt} shows the separation of the $\pi^{+}$ from the proton and kaon tracks.

For the $\pi^{-}$ tracks it was necessary to impose additional cuts to remove electron contamination. Cuts opposite to those used in the electron identification using the EC inner energy and number of photoelectrons in the CC are used.

\subsection{Neutral pion identification}

Neutral pions are identified via their primary decay mode, $\pi^{0} \rightarrow \gamma \gamma$. Both photons are detected in the CLAS EC, and the invariant mass of the photon pair was computed to reconstruct the $\pi^{0}$.

\begin{figure}
\centering
\includegraphics[width=7.5cm]{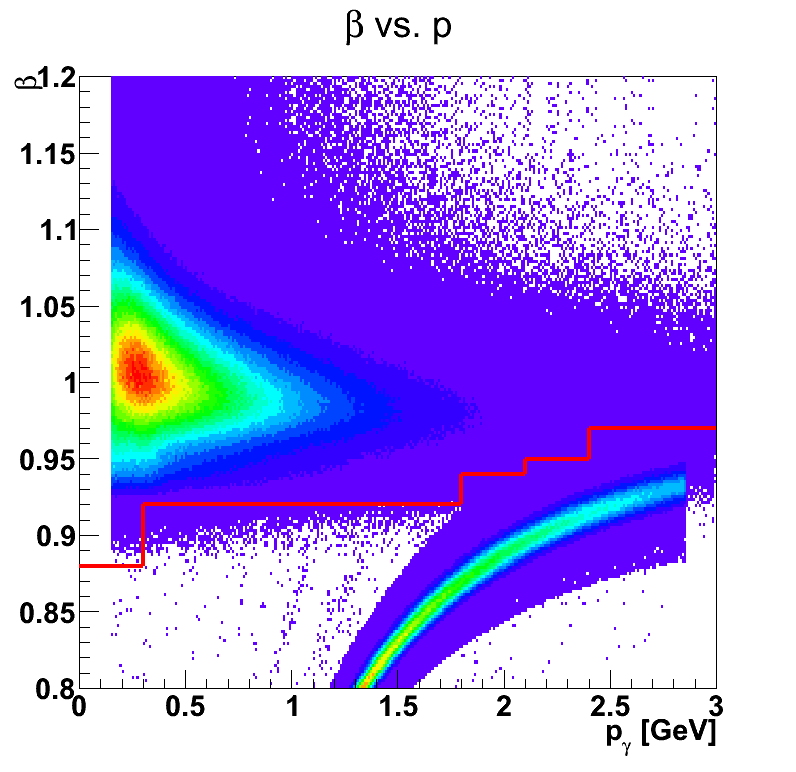}
\caption{(Color Online) $\beta$ vs momentum for hits in the CLAS calorimeter. A momentum-dependent cut, shown by the solid red line, was used to separate photons from neutrons.}
\label{fig:betap}
\end{figure}

Photons are distinguished from neutrons by measuring the $\beta$ of each hit in the calorimeter, and performing a momentum dependent cut, as shown in Fig.~\ref{fig:betap}, where $\beta$ was measured based on the EC hit-time as 

\begin{equation}
\beta = \frac{L_{\mbox{EC}}/c}{t_{\mbox{EC}}-t_{\mbox{start}}},
\end{equation}
where $L_{\mbox{EC}}$ is the distance from the vertex to the EC hit, $t_{\mbox{EC}}$ is the time measured by the calorimeter, $t_{\mbox{start}}$ is the start time of the event with respect to the machine radio frequency, and $c$ is the speed of light. 
A cut is necessary to remove events in which the angle between the two photons is smaller than the resolution of the EC, leading to an inaccurate reconstruction of the invariant mass, and another is made on the hit position in the calorimeter, which is similar to that used in the electron identification.

\begin{figure}
\centering
\includegraphics[width=8cm]{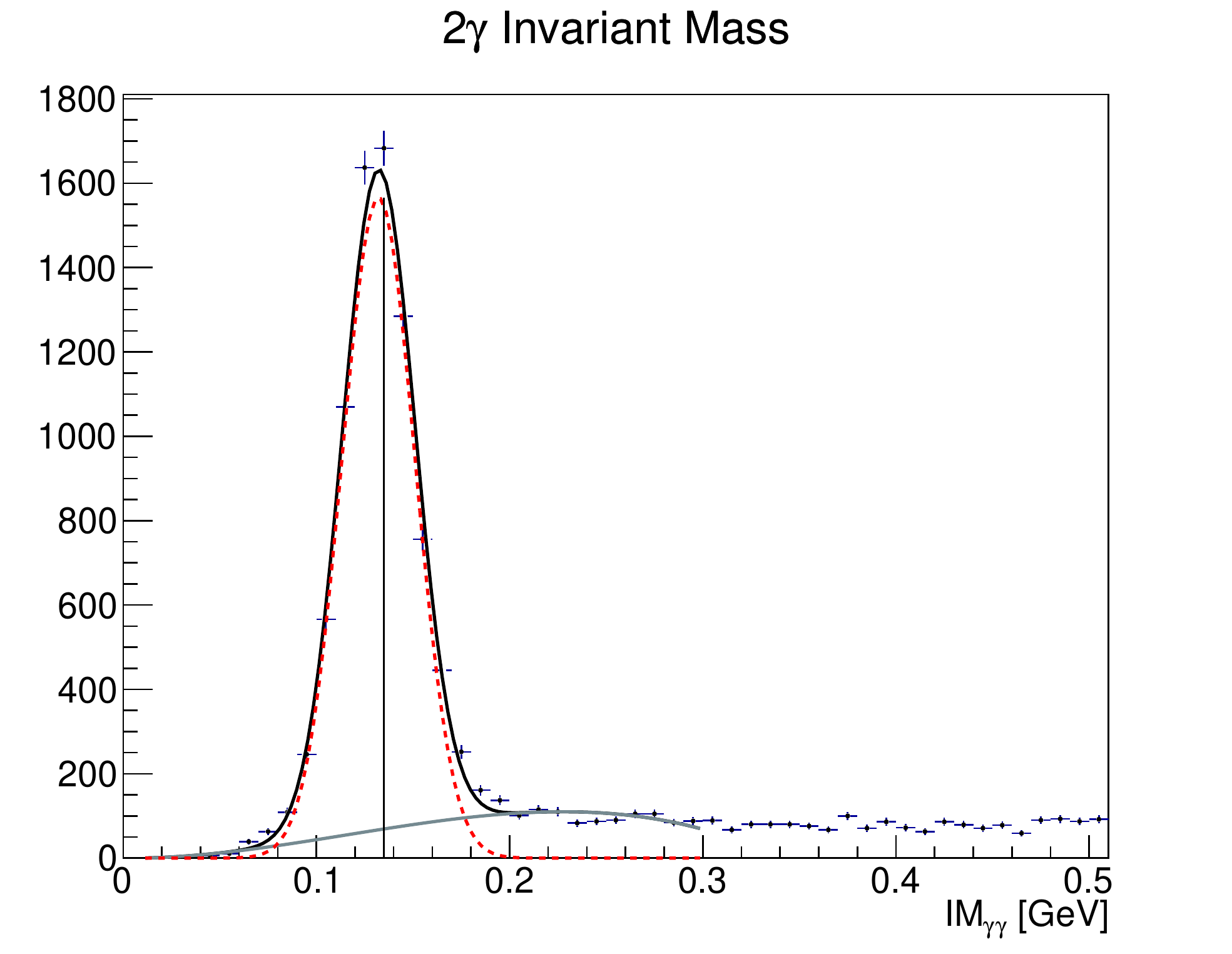}
\caption{(Color Online) Invariant mass of two photons in a single bin in $x_B$, $P_{T}$, and $\phi$ for the positive helicity state. The invariant mass was fit with a Gaussian (red dotted curve) plus a polynomial background (grey curve), which is integrated to determine the number of neutral pions in the bin.}
\label{fig:imgg}
\end{figure}

For events with the number of identified photons, $N_{\gamma} \ge 2$, the invariant mass of each potential photon pair was binned by helicity, $\phi$, and the remaining kinematic variables. Each bin was then fit with a Gaussian plus polynomial background to determine the number of events in each bin, as shown in Fig.~\ref{fig:imgg}. The background polynomial was of first, second, or third order depending on the shape of the background in that particular kinematic bin. It was ensured that the background function was the same for each helicity state in order to decrease the likelihood of an asymmetry arising due to differences in the background function. The number of (background subtracted) events in each kinematic bin were determined by integrating the Gaussian signal function, $f(M_{\gamma\gamma})$, over $\pm3\sigma$, where $\sigma$ was the standard deviation of the Gaussian in that bin, and computing the number of events as shown in Eq.~\ref{eq:histint}.

\begin{equation}
N_{events}=\frac{1}{\mbox{bin size}}\int_{-3\sigma}^{+3\sigma} \! f(M_{\gamma\gamma}) \, \mathrm{d} M_{\gamma\gamma}
\label{eq:histint}
\end{equation}

\subsection{Kinematic coverage}

The kinematic requirements for DIS are $W>2$ GeV and $Q^2>1$ GeV$^2$, where $W$ is the invariant mass of the final state and $Q^2$ is the virtuality of the exchange photon, or four-momentum transfer squared, from the incident lepton to the target, given by $W^2=( P + q)^2$ and $Q^2 = -q^2 = -(k-k')^2$. Here $P$ is the 4-vector of the target and $k(k')$ is the 4-vector of the incoming (outgoing) lepton.

The measurements were performed in terms of the kinematic variables $x_B$, $z$, $P_{T}$, and $Q^{2}$. Here, $x_B$ is the momentum fraction carried by the quark in the proton, which is defined by $x=\frac{Q^2}{2P \cdot q}$. The variable $z$ is the momentum fraction carried away by the produced hadron, which is defined manifestly as a Lorentz scalar by $z=\frac{P \cdot P_h}{P \cdot q}$, and $P_{T}$ is the transverse momentum of the outgoing hadron relative to $\mathbf q$.

\begin{figure}[!bh]
\centering
\includegraphics[width=6.5cm]{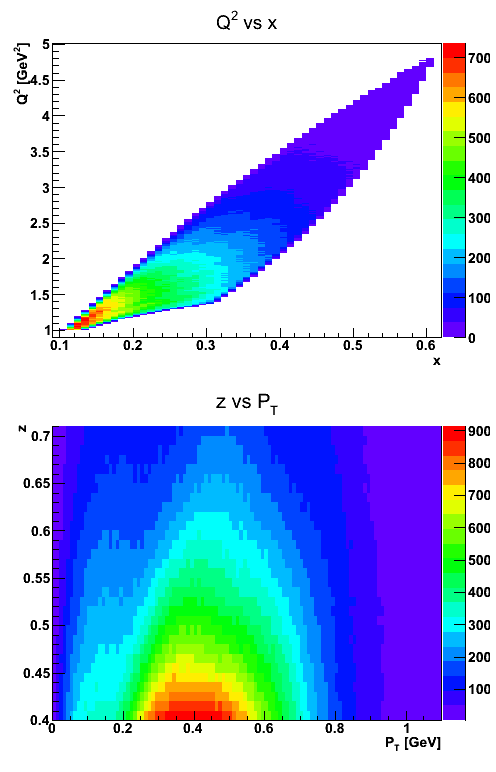}
\caption{(Color Online) Kinematic coverage for SIDIS events. $A_{LU}^{\sin\phi}$ was binned in $z$, $x_B$, $P_{T}$, and $Q^{2}$, so the upper panel shows $Q^{2}$ vs. $x_B$ and the lower panel shows $z$ vs. $P_{T}$, where $z$ has been constrained to the SIDIS region of $0.4<z<0.7$. The distributions are shown for $\pi^{+}$ events, but the kinematic coverage for $\pi^{-}$ and $\pi^{0}$ is similar.}
\label{fig:kin}
\end{figure}

One advantage of performing the experiment in CLAS was the wide kinematic coverage available for analysis. Coverage in $x_B$ was available from 0.1 to 0.6, corresponding to a $Q^{2}$ range from 1.0 GeV$^{2}$ to 4.5 GeV$^{2}$. Coverage in $W$ extends from our imposed minimum of 2 GeV to 3.1 GeV, and $P_{T}$ extends from 0 to 1 GeV. A cut was imposed on $0.4<z<0.7$ to select semi-inclusive events. Events with missing mass $M_{X}$ below 1.2 GeV are removed, which in conjunction with the cut on z effectively removed contamination from exclusive events such as $ep \rightarrow e\pi^{+}n$.  The full range of kinematic coverage for this experiment is shown in Fig.~\ref{fig:kin}.

\section{Data Analysis}

The data were analyzed by computing beam-spin asymmetries (BSAs), as shown in Eq.~\ref{eq:bsa} in each kinematic bin. Here $N^{+}$ is the number of events from the positive helicity beam, $N^{-}$ is the number of events with the negative helicity beam, and $P_{e}$ is the average beam polarization.

\begin{equation}
A(\phi) = \frac{1}{P_{e}}\frac{N^{+}-N^{-}}{N^{+}+N^{-}}
\label{eq:bsa}
\end{equation}

\noindent Statistical uncertainties were computed for each BSA for charged pions, as 

\begin{equation}
\delta A = \frac{1}{P_{e}}\sqrt{ \frac{ 1-(P_{e}A)^{2}}{N^{+} + N^{-}}},
\label{eq:ebsa}
\end{equation}

\noindent which was computed from the standard uncertainty on the number of events in each helicity state, $\delta N^{\pm} = \sqrt{N^{\pm}}$. Uncertainties for $\pi^{0}$ BSAs include an additional factor to account for the background subtraction that slightly increases the uncertainty. The uncertainty on the beam polarization from the multiple M$\o$ller measurements was included in the systematic uncertainty.

The BSA was binned in ten bins in $z$ from 0 to 1 in order to see the $z$-dependence, but when not looking at the $z$ dependence, cuts were applied to keep only $0.4<z<0.7$ to limit results only to the SIDIS kinematic region. After the z cut, the data were binned in five bins in $x_B$ from 0.1 to 0.6, five bins in $P_{T}$ from 0 to 1 GeV, and five bins in $Q^{2}$ from 1 to 4.5 GeV$^{2}$. The data were also analyzed using two-dimensional binning in $x_B$ and $P_{T}$ over the same ranges, since the accessible TMDs are functions of $x_B$ and $k_{T}$ (which can be approximately related to $P_{T}$~\cite{Schweitzer:2010tt}). 

In each kinematic bin, the BSAs were binned in $\phi$ and fit to determine $A_{LU}^{\sin\phi}$. The fitting function, $f(\phi)$, was derived from Eq.~\ref{eq:cs} as

\begin{equation}
f(\phi) = \frac{A \sin\phi}{1+B\cos\phi+C\cos2\phi},
\label{eq:fit}
\end{equation}
and the coefficient $A$ was extracted as the value of $A_{LU}^{\sin\phi}$ in each bin. An example of the fitting procedure is shown in Fig.~\ref{fig:rand}. The systematic uncertainty due to the fitting process was checked by fitting with other functions such as $A\sin\phi$ or $A\sin\phi + B\sin2\phi$. It should be noted that the coefficients on $\sin2\phi$ were found to be consistent with zero. 

\begin{figure}
\centering
\includegraphics[width=7cm]{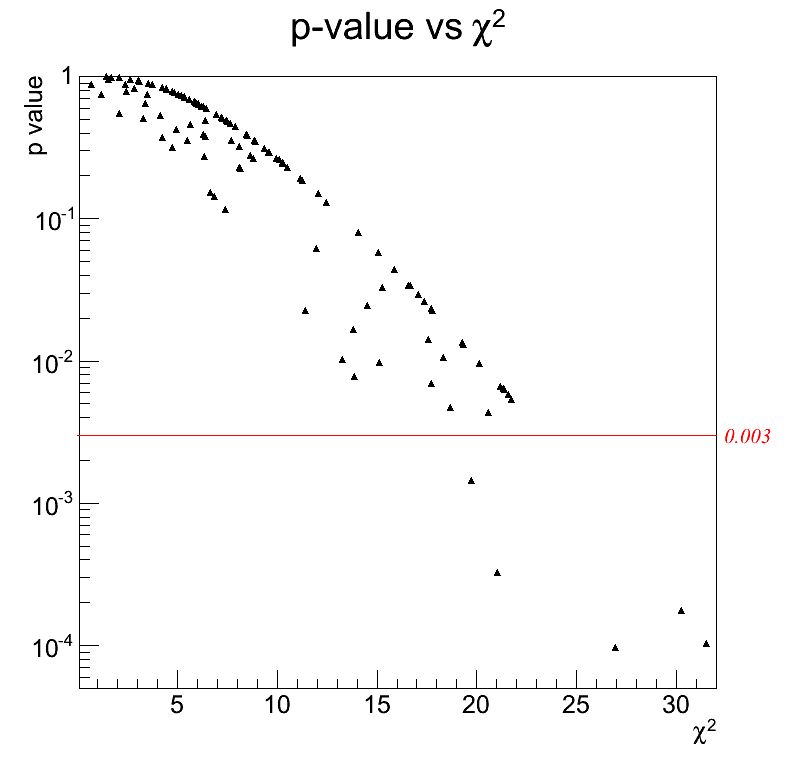}
\caption{$p$-value vs. $\chi^{2}$. Fits for which the $p$-value $<$ 0.003 did not pass our hypothesis test and were removed.}
\label{fig:pval}
\end{figure}

All fits were performed using a $\chi^{2}$ minimization with the TMinuit class in ROOT~\cite{ROOT}. Uncertainties on the fit coefficients were computed in MINUIT from the $\chi^{2}$ of each fit. The goodness-of-fit was evaluated by hypothesis testing. The null hypothesis $H_{0}$ was defined by the statement that the BSA distributions were consistent with our fit function. A significance level was set to 0.003, so if the fit in a particular bin returns a $p$-value less than 0.003, the hypothesis was rejected, and that fit was removed from the analysis. A plot of the $p$-values for each fit vs. $\chi^{2}$ is shown in Fig.~\ref{fig:pval}, where the red line indicates our significance level. The removal of fits with a very low $p$-value also served to remove fits with a $\chi^{2}$ higher than would be expected based on our statistics. The measured $\chi^{2}$ distribution was compared to the expected $\chi^{2}$ probability density function (p.d.f.) computed for eight degrees of freedom and normalized to our bin size and number of entries, as given by Eq.~\ref{eq:chi2}

\begin{equation}
f(x) = \frac{2\times118}{96}x^{3}e^{-x/2}.
\label{eq:chi2}
\end{equation}

\begin{figure}
\centering
\includegraphics[width=7cm]{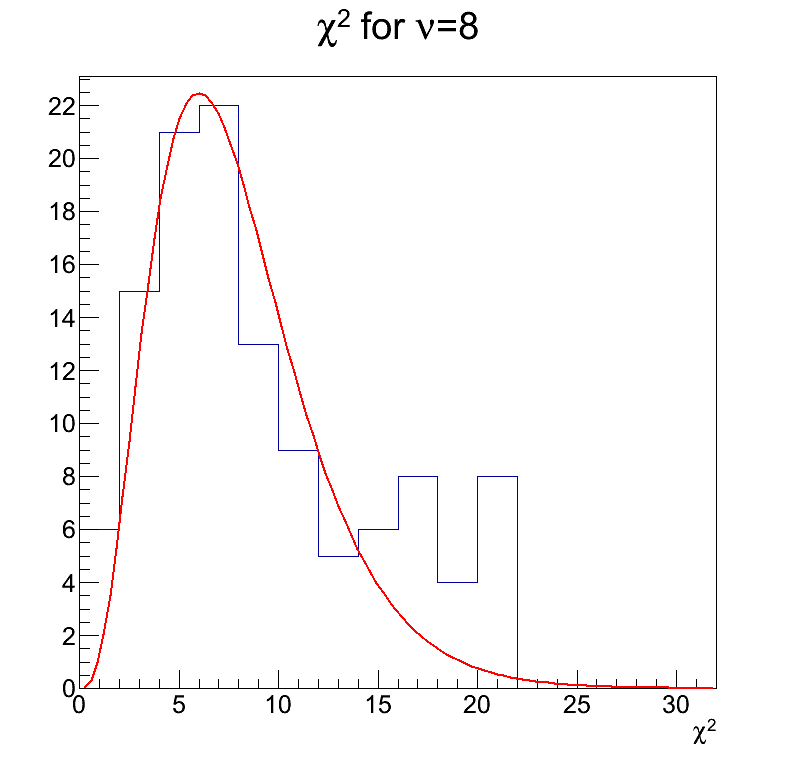}
\caption{Comparison between the measured (blue) and expected (red) $\chi^{2}$ distributions.}
\label{fig:chi2}
\end{figure}

\noindent The p.d.f. is written as $f(x)$, where $x_B$ is the $\chi^{2}$. By definition, the mean $\nu$ of the distribution is eight for the degrees of freedom, and the leading coefficient is given by $2^{-\nu/2}/\Gamma(\nu/2) = 1/96$. The distribution was then scaled by a factor of $2\times118$ to match the number of entries and bin size of the measured $\chi^{2}$ histogram. The plot of $\chi^{2}$ is shown in Fig.~\ref{fig:chi2}, and it can be seen that the measured $\chi^{2}$ distribution closely matches the expected distribution. The reduced $\chi^2$ between the two distributions is 1.37.

\section{Systematic Studies}

The systematic uncertainty was estimated to be smaller than or nearly equal to the statistical uncertainty in all measured kinematics. These estimations were carried out by varying the analytical techniques and measuring the resulting change in $A_{LU}^{\sin\phi}$. The total systematic uncertainty due to these variations was estimated to be 0.006 for $\pi^{+}$, 0.007 for $\pi^{-}$, and 0.009 for $\pi^{0}$. Table~\ref{tab:sys} gives the systematic uncertainty due to each variation for each pion channel. Contributions were estimated from acceptance effects and pion contamination to the identified electrons. The analysis was repeated using randomly generated helicity states, which gave an asymmetry consistent with zero. 

\begin{table}
\centering
\vspace{0.3cm}
\begin{tabular}{|c|c|c|c|}
\hline
Sources of uncertainty  & \multicolumn{3}{c|}{Average uncertainty}\\
\cline{2-4}
 &$\pi^{+}$ & $\pi^{-}$ & $\pi^{0}$ \\
\hline
EC $E_{inner}$ cut &  0.0017& 0.0030 & 0.0003\\
EC sampling fraction &  0.0005 & 0.0016 & 0.0020\\
Electron fiducial cut &  0.0011 & 0.0029 & 0.0020\\
Vertex cut &  0.0021 & 0.0029 & 0.0036\\
Pion identification &  0.0007 & 0.0028 & - \\
Pion fiducial cut &  0.0018 & 0.0040 & -\\
Missing mass cut & 0.0052 & 0.0029 & 0.0064\\
Background subtraction &  -& -& 0.005 \\
Background asymmetry &  -&- & 0.007\\
Fitting function &   0.0007& 0.0011 & 0.0010\\
Beam polarization &   0.0004 & 0.0006 & 0.0008\\
\hline
Total &  0.006 & 0.008 & 0.012\\
\hline
Statistical uncertainty  & 0.005 & 0.014 & 0.012\\
\hline
\end{tabular}
\caption{Sources systematic uncertainty. The second column gives the average absolute uncertainty from each source. For comparison, the average statistical uncertainty is given.}\label{tab:sys}
\end{table}

The inclusive beam-charge asymmetry (BCA) was also calculated on a run-by-run basis to insure that the integrated number events for each helicity state remained constant over the entire run period. The integrated BCA was measured to be 0.003. To test the impact of this asymmetry on the measured physics asymmetry, $A_{LU}^{\sin\phi}$ was calculated independently for all events from runs with a positive BCA and again for all events from runs with a negative BCA. The two values were found to be identical to a precision much smaller than the quoted systematic uncertainties. It is hence concluded that the BCA does not contribute to the measured physics asymmetry.

\subsection{Variation of particle identification}

Particle identification cuts were varied for both electron and pion identification routines. The electron identification was tested by varying the minimum and maximum values for the sampling fraction, EC $E_{\mbox{inner}}$, fiducial, and vertex cuts. The charged pion identification was tested by switching between the nominal cut on $\Delta t$ and an analogous cut on the $\beta = v/c$ for each track, as well as variations of the pion fiducial cuts. $\pi^{0}$ identification was tested by modifying the background subtraction technique, and by computing the asymmetry of the background itself.

A significant contamination of pions in the electron sample could give a large contribution to the systematic uncertainty. The electron identification cut most sensitive to pion contamination was that on the EC inner energy, nominally at $E_{\mbox{inner}}>55$ MeV, which is equivalent to 3$\sigma$ from the pion peak. To test for pion contamination, the data were analyzed comparing the cut at $E_{\mbox{inner}}>0$ MeV, giving maximum pion contamination, and $E_{\mbox{inner}}>100$ MeV, cutting well into the sample of good electrons, and hence removing any trace of contamination due to misidentified $\pi^{-}$. The difference in $A_{LU}^{\sin\phi}$ due to this variation was negligible in comparison with the other sources of uncertainty.

 As an additional check on the possibility of pion contamination, the EC $E_{\mbox{inner}}$ distribution was generated for five separate ranges of both $Q^{2}$ and $W$, and then fit with a functional form that combined a Gaussian with a polynomial. The two functions were integrated individually, and the percentage of events passing the electron identification that were actually misidentified pions was computed to be less than $10^{-3}$ in every bin.

The uncertainty due to the value of the missing mass cut was computed by comparing the nominal results, which use a missing mass cut at 1.2 GeV, with results stemming from missing mass cuts at 1.1 GeV and 1.3 GeV. 

\subsection{Uncertainty due to fitting procedure}

Because $A_{LU}^{\sin\phi}$ was extracted as the coefficient of a fit, it is important that the fitting technique be very reliable. This was tested by variation of the fitting procedure, fitting of data using a random beam helicity, and fitting of simulated data seeded with a known asymmetry. 

The nominal fitting function shown in Eq.~\ref{eq:fit} was derived from the SIDIS cross section as shown in Eq.~\ref{eq:cs},  so this function gives the most physically accurate description of the beam-spin asymmetry. The $B$ and $C$ coefficients result from an unstable equilibrium in the $\chi^2$ minimization, so small changes in the data can cause large fluctuations in these values. We did not use this fit to make a measurement of those values, but their inclusion still made the shape of the fitting function more realistic.

To determine the systematic uncertainty due to this fit, the BSAs were fit with a simplified version of the fitting function, $A\sin\phi$, and the resulting $A$ coefficients were compared to those from the full fit to compute a value of systematic uncertainty. Other functions were tested as well, including $\frac{A\sin\phi}{1+B\cos\phi}$, $\frac{A\sin\phi}{1+B\cos2\phi}$, and $A\sin\phi+B\sin2\phi$. The first two of these were used to determine the contribution of the cosine term to the fit, and the third tests for a possible contribution from higher harmonics, which were observed to be consistent with zero. The variation of the $A$ coefficient due to fits with this variety of functions is seen be only about 13\% of the total systematic uncertainty, as shown in Table~\ref{tab:sys}.

\begin{figure}
\includegraphics[width=8cm]{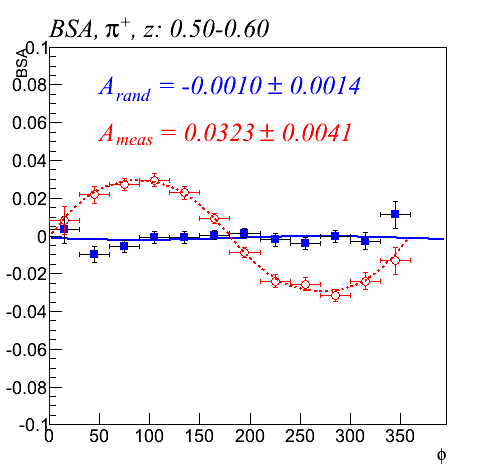}
\caption{(Color Online) Beam-spin asymmetries for one bin in $z$, plotted in red, open symbols (blue, solid symbols) using measured (random) helicities. Both are fit using Eq.~\ref{eq:fit}. An estimate of systematic uncertainty was made by fitting the BSA using a random helicity, which is expected to be zero, and comparing the result with the null hypothesis.}
\label{fig:rand}
\end{figure}

Another method to test the fitting procedure is to artificially modify the data to produce a known result, and then measure the discrepancy between the expected result and that provided by the fit. This was done in two ways. The first was to seed the data with a random helicity, resulting in an asymmetry of zero, as shown in Fig.~\ref{fig:rand}. The second was to test the fitting procedure on simulated data with a known asymmetry. In both cases the fits yield values within 1.5$\sigma$ of the expected value. 

\subsection{Uncertainty due to beam polarization}


Any uncertainty in the measurement of beam polarization will lead to an uncertainty in $A_{LU}^{\sin\phi}$. The beam polarization was measured periodically during the run using a M$\o$ller polarimeter, with an average measurement of $P_{e}=0.751$. Known systematic effects provide a $\delta P_{e}$ of 3\% (relative)~\cite{clasnim}, which was then used to compute the contribution of the beam polarization to the uncertainty on $A_{LU}^{\sin\phi}$, as shown in Eq.~\ref{eq:polerr}. 
\begin{equation}
\delta A_{LU} = \frac{\delta P_{e}}{P_{e}}A_{LU}
\label{eq:polerr}
\end{equation}
These values were averaged over all bins to give the value in Table~\ref{tab:sys}.

\subsection{Acceptance effects}

The effect of acceptance on beam-spin asymmetries was found to be negligible in this analysis. The CLAS data can be susceptible to effects from detector inefficiencies, but these effects cancel out for BSAs as long as the acceptance is the same for positive and negative beam helicity and the bin size used in the analysis is sufficiently small.

The CLAS acceptance was computed using a Monte Carlo simulation. Data were generated using a Lepto-based event generator with realistic physics for semi-inclusive DIS~\cite{Ingelman:1996mq}, and the Monte Carlo was performed using a GEANT-based detector simulation~\cite{Brun:1978fy}. Each kinematic bin was analyzed using the raw generated data first, and then again using the data that had passed through the simulated CLAS detector. The efficiency in each kinematic bin was computed as the ratio of the number of events detected in that bin after the Geant detector simulation to the number of events seen in that bin from the generated data sample, 

\begin{equation}
A = \frac{ N_{R}}{N_{G}},
\label{eq:anrng}
\end{equation}

\noindent
 ($R$=reconstructed and $G$=generated). The experimental data were then corrected in each bin using this calculated efficiency, 

\begin{equation}
N' = \frac{N_{M}}{A},
\label{eq:nnma}
\end{equation}

\noindent
($N'$ is corrected and $M$=measured). The acceptance was computed using only events with positive helicity and only events with negative helicity, and the ratio of the two were checked in every bin to insure they were in agreement with unity to better than 1$\sigma$. Since the CLAS acceptance was seen to be equal for each helicity state, the corrected $BSA$, which we call $BSA'$ is unchanged by the acceptance correction, as demonstrated in Eq.~\ref{eq:bsaacc}.

\begin{equation}
BSA' = \frac{N'^{+}-N'^{-}}{N'^{+}-N'^{-}} = \frac{N^{+}/A^{+} - N^{-}/A^{-}}{N^{+}/A^{+}+N^{-}/A^{-}} 
\label{eq:bsaacc}
\end{equation}

\noindent
Then, if $A^{+}=A^{-}=A$,

\begin{equation}
BSA' = \frac{N^{+}/A-N^{-}/A}{N^{+}/A+N^{-}/A} = \frac{N^{+}-N^{-}}{N^{+}+N^{-}} = BSA.
\label{eq:bsaacc2}
\end{equation}

\section{Results}

$A_{LU}^{\sin\phi}$ has been measured with good statistics in all three pion channels. Data were binned using five bins in $x_B$ from 0.1 to 0.6, five bins in $P_{T}$ from 0 to 1 GeV, five bins in $Q^{2}$ from 1.0 to 4.5 GeV$^2$, and ten bins in $z$ from 0 to 1, though when looking at dependencies on $x_B$, $P_{T}$, or $Q^{2}$, only the $z$ bins between 0.4 and 0.7 were used. 

\begin{figure*}
\centering
\includegraphics[width=16cm]{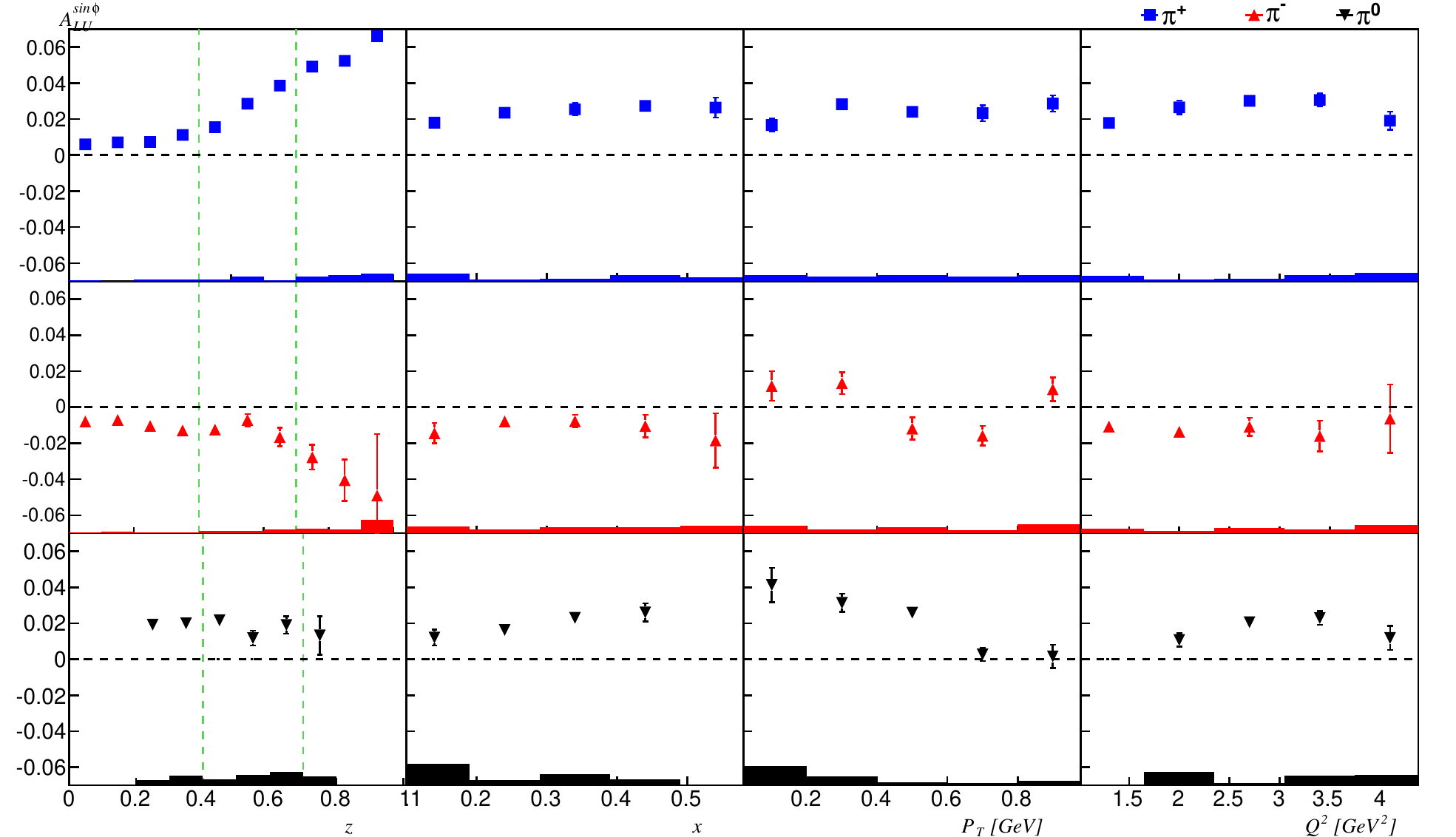}
\caption{(Color online) $A_{LU}^{\sin\phi}$ vs. $z$, $x_B$, $P_{T}$, and $Q^{2}$ after integration over other the other kinematic variables for each pion channel. The integration range in $z$ for SIDIS kinematics is for $0.4<z<0.7$. The error bars represent statistical uncertainties and the shaded regions represent the systematic uncertainties. The top row shows $\pi^{+}$, the center row shows $\pi^{-}$, and the bottom row shows $\pi^{0}$.}
\label{fig:alu1d}
\end{figure*}

\begin{figure*}
\centering
\includegraphics[width=16cm]{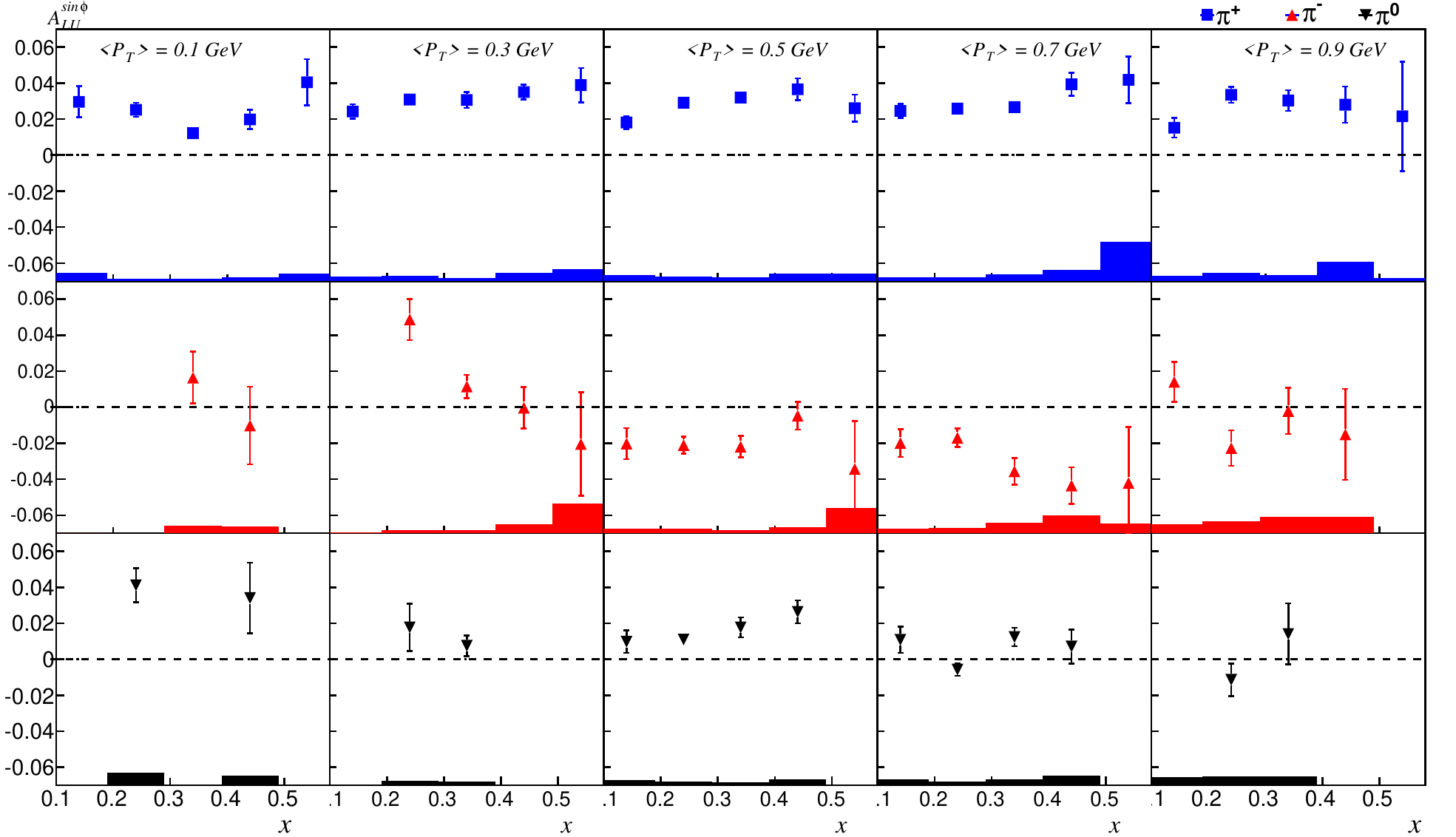}
\caption{(Color online) $A_{LU}^{\sin\phi}$ vs $x_B$ in different $P_{T}$ bins. The error bars represent statistical uncertainties and the shaded regions at the bottom represent systematic uncertainties. The top row is $\pi^{+}$, the middle is $\pi^{-}$, and the bottom $\pi^{0}$. Each column represents a different bin in $P_{T}$. The results are integrated over all $Q^{2}$ and $0.4<z<0.7$.}
\label{fig:alu2d}
\end{figure*}

Fig.~\ref{fig:alu1d} and Table~\ref{tab:alu1d} show $A_{LU}^{\sin\phi}$ in one dimension vs $x_B$, $z$, $P_{T}$, and $Q^{2}$. The full data set is contained in the CLAS database ~\cite{clasdb}. For each of these plots, all kinematics other than the demonstrated dependence were integrated. When integrating, $x_B$, $z$, and $P_{T}$ were integrated over their entire range, and $z$ were integrated from 0.4 to 0.7. Fig.~\ref{fig:alu2d} and Table~\ref{tab:alu2d} show $A_{LU}^{\sin\phi}$ in two dimensions of $x_B$ and $P_{T}$. Here we have integrated over all $Q^{2}$ bins and $0.4<z<0.7$. 

\begin{figure*}
\centering
\includegraphics[width=17.5cm]{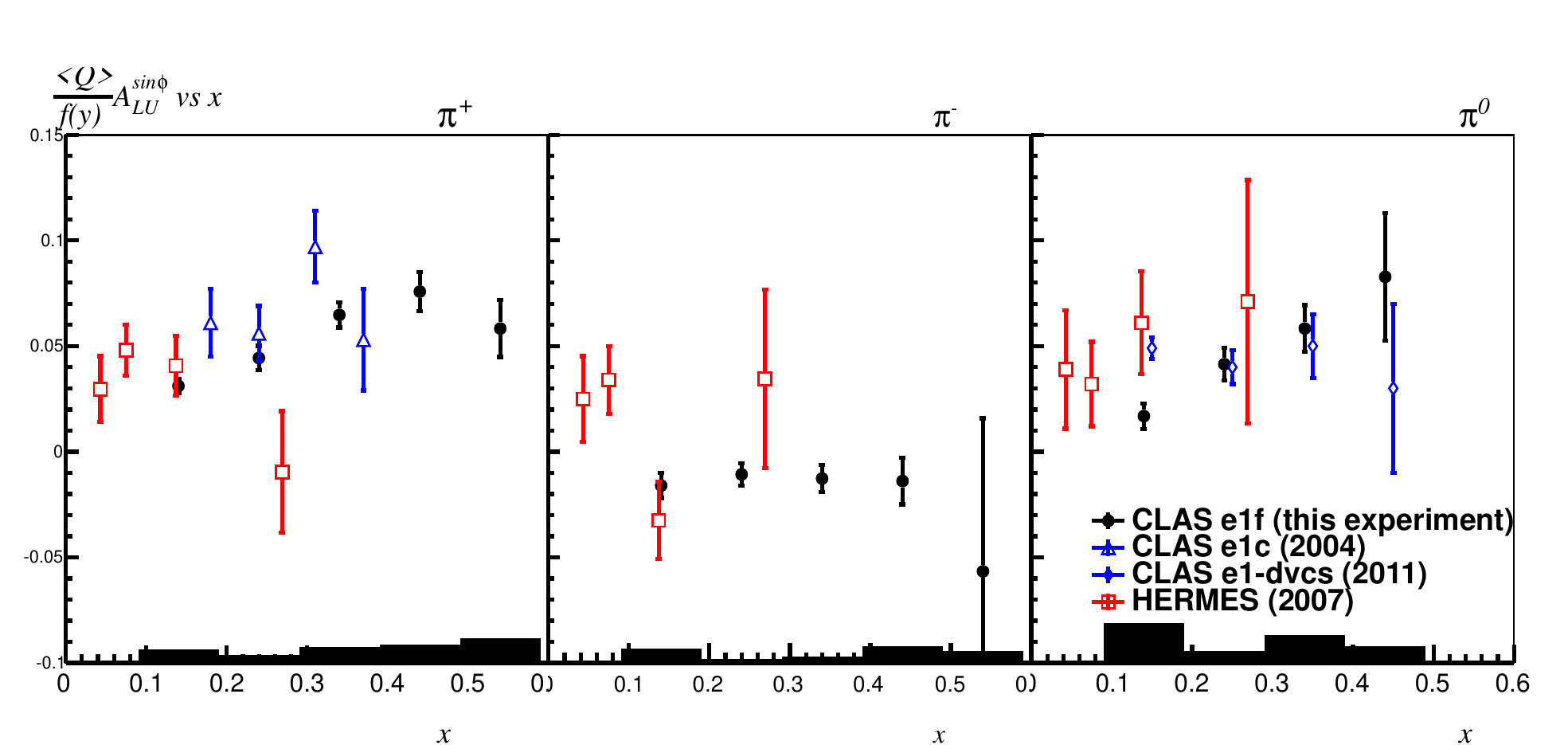}
\caption{(Color online) Comparison of $A_{LU}^{\sin\phi}$ vs $x_B$ between several datasets, each scaled by a factor of $<Q>/f(y)$, where $f(y)$ is given by Eq.~\ref{eq:fy}. The left panel is $\pi^{+}$, the center is $\pi^{-}$, and the right is $\pi^{0}$. Here the solid (black) circles are this experiment, the open (red) squares are from HERMES~\cite{Airapetian:2006rx}, and the open (blue) triangles are from previous CLAS publications~\cite{e16, Avakian:2003pk, Aghasyan:2011ha}. The error bars are statistical uncertainty and the solid bands show the systematic uncertainty for the current experiment.}
\label{fig:herm2}
\end{figure*}

The observable $A^{\sin\phi}_{LU}$ has been previously measured for $\pi^+$ and $\pi^0$ using CLAS and for all three pion channels at HERMES. Fig.~\ref{fig:herm2} shows the comparison between the current experiment and the previously published data. The previously published CLAS $\pi^0$ measurement (the E1-dvcs run) utilized an inner calorimeter that greatly increased photon detection at small angles. With that inner calorimeter, the E1-dvcs experiment had more uniform coverage in $\phi$ for $\pi^0$s at low $x_B$. This inner calorimeter was not available at the time of the E1-f run period from which the current data were obtained, and, as a result, the systematic uncertainties in the value for $A^{\sin\phi}_{LU} $ obtained at the lowest $x_B$ value for the experiment reported here are considerably larger than those for the previous CLAS measurement. This large uncertainty for the lowest $x_B$ point obtained here prevents rejection of the null hypothesis that there is no difference between the two CLAS measurements ($p = 0.885$).

Because HERMES results were obtained at a higher beam energy than the CLAS, it is necessary to scale the data by a factor of $<Q>/f(y)$ to make a valid comparison, where $f(y)$ is given by
\begin{equation}
f(y) \approx \frac{y\sqrt{1-y}}{1-y+y^{2}/2}
\label{eq:fy}
\end{equation}
\noindent
and $y$ is again the energy fraction~\cite{Bacchetta:2006tn}.

It has been observed in measurements at Belle that the Collins mechanism results in opposite $z$-dependence in the two charged pion channels~\cite{Seidl:2008xc}, so in the case that the asymmetry is dominated by the Collins type contribution, $\pi^{+}$ and $\pi^{-}$ will have  opposite sign for the $\sin\phi$ moments.  

It is expected that $\pi^{0}$s will give the same sign of asymmetry as $\pi^{+}$. Isospin symmetry predicts that the magnitude of $A_{LU}^{\sin\phi}$ for $\pi^{0}$s should give a weighted average of the moments from $\pi^{+}$ and $\pi^{-}$. Since the Collins contribution to $\pi^{+}$ and $\pi^{-}$ is roughly equal and opposite, one would expect a very small asymmetry for $\pi^{0}$ from this effect. Since it is very far from zero, one could argue that other contributions are relevant for $\pi^{0}$. 

There has been a phenomenological work that attempted to extract twist­3 functions from the existing data ~\cite{Efremov:2002sd}, however higher precision data are needed for model independent studies of TMDs~\cite{Avakian:2009jt, Avakian:2010br, Efremov:2006qm, Pasquini:2011tk}. The current measurement provides a significant upgrade of the previous CLAS~\cite{Avakian:2003pk} and HERMES~\cite{Airapetian:2006rx} results, which should be sufficient to contribute to an updated analysis of the relevant twist-3 TMDs, with the newly added benefit of viewing flavor separation with the large improvement in the $\pi^{-}$ measurement.

\begin{figure*}
\centering
\includegraphics[width=17.5cm]{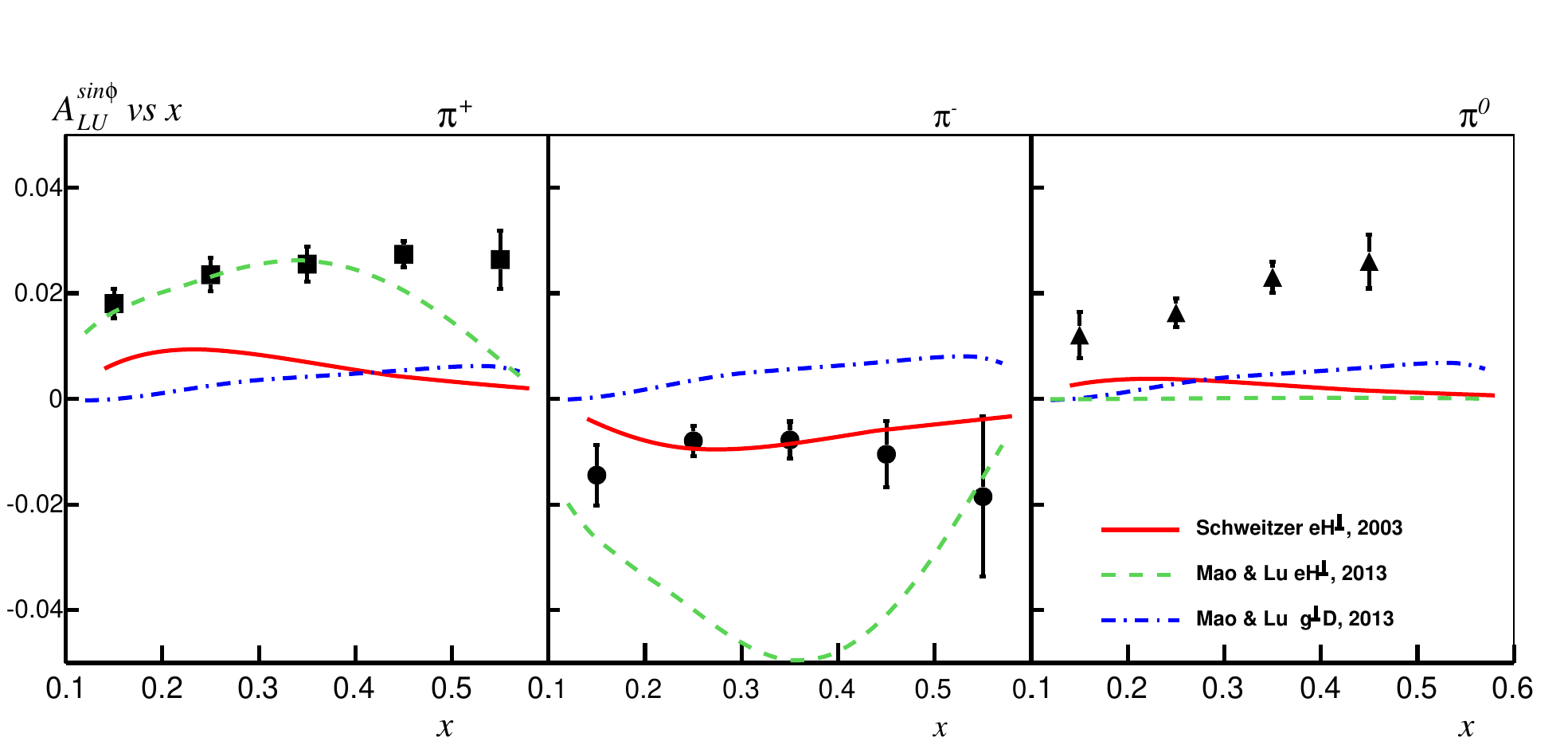}
\caption{(Color online) Comparison of measurement to theoretical models taking into account either the contributions due to the $e \otimes H_{1}^{\perp}$ term as calculated in~\cite{Schweitzer:2003uy} (solid curves, red) or as calculated in~\cite{Mao:2012dk, Mao:2013waa} (dashed, green). The $g^{\perp} \otimes D_{1}$ contribution (dashed-dot, blue curves) ~\cite{Mao:2012dk, Mao:2013waa} are also shown, which were calculated utilizing the spectator model described in~\cite{Bacchetta:2003rz}.}
\label{fig:model}
\end{figure*}

In Fig.~\ref{fig:model} our results are compared to a model described in~\cite{Efremov:2002ut,Schweitzer:2003uy}, which takes into account only the contribution of the $e(x) \otimes H_1^{\perp}$ term to the $\sin\phi$ moment, where prescription from~\cite{Efremov:2006qm} is used to model the Collins contribution, as well as another model as described in~\cite{Mao:2012dk, Mao:2013waa}, which also accounts for the $g^{\perp} \otimes D_{1}$ contribution using two different spectator models~\cite{Bacchetta:2008af,Bacchetta:2003rz}. The model predictions are computed specifically for the E1-f kinematics. The opposite sign of the two charged pion channels is consistent with a significant contribution related to the Collins function~\cite{Schweitzer:2003uy}, but the difference in scale for $\pi^{+}$ and $\pi^{0}$ in particular suggests that the other three contributions to the structure function must also play relevant roles. The model by Mao and Lu, when using the spectator model in~\cite{Bacchetta:2003rz}, describes our data well for $\pi^{+}$ and accurately predicts the sign of the asymmetry for $\pi^{-}$ (though the predicted magnitude for $\pi^{-}$­ is much larger than the present data)", while showing a very small contribution for $\pi^{0}$. The prediction by Mao and Lu using the spectator model described in~\cite{Bacchetta:2008af}, however, is highly inconsistent with our measurement, especially for $\pi^{-}$, where the prediction is large and of opposite sign. 

The comparison of $A_{LU}^{\sin\phi}$ measurements for all 3 pions (Fig.~\ref{fig:model}), with contributions from only the Collins effect, indicates that Sivers-type contributions $g^\perp\otimes D_1$\cite{Mao:2012dk} may be significant for $\pi^+$ and $\pi^0$ but
small for $\pi^-$. This is consistent with the latest observations by HERMES and COMPASS~\cite{Airapetian:2009ti,Adolph:2012sp,Adolph:2012sn}, where a large Collins effect 
was observed for charged pions, while the Sivers effect was found to be
significant only for  $\pi^+$.
A complete description would be the sum of these two contributions, as well as the contributions from the two twist-3 fragmentation functions.

\begin{table*}
\centering
\vspace{0.3cm}
\begin{tabular}{|c|c|c|c|c|c|c|c|}
\hline
$<z>$ & $<x>$ & $<P_{T}>$ & $<Q^{2}>$ & $<y>$ & $A_{LU}^{\sin\phi}$, $\pi^{+}$ $\pm (stat) \pm (syst)$ & $A_{LU}^{\sin\phi}$, $\pi^{-}$ $\pm (stat) \pm (syst)$& $A_{LU}^{\sin\phi}$, $\pi^{0}$ $\pm (stat) \pm (syst)$\\
\hline
\bf{ 0.05} & 0.24 &  0.18  &  1.80   &0.72& 0.0059 $\pm$ 0.0004 $\pm$ 0.0004& -0.008 $\pm$ 0.001 $\pm$ 0.000& \\
\bf{0.15} & 0.26 &  0.29 & 1.84  &0.71& 0.0071 $\pm$ 0.0009 $\pm$ 0.0002& -0.007 $\pm$ 0.002 $\pm$ 0.001& \\
\bf{0.25} & 0.27 & 0.37 & 1.87 &0.70& 0.007 $\pm$ 0.001 $\pm$ 0.001 & -0.010 $\pm$ 0.001 $\pm$ 0.000 & 0.019 $\pm$ 0.002 $\pm$ 0.003\\
\bf{0.35} & 0.27 & 0.42 & 1.90 &0.69& 0.011 $\pm$ 0.001 $\pm$ 0.001 & -0.013 $\pm$ 0.002 $\pm$ 0.000 & 0.020 $\pm$ 0.002 $\pm$ 0.005\\
\bf{0.45} & 0.28 & 0.45 & 1.92 &0.68& 0.015 $\pm$ 0.001  $\pm$ 0.001 & -0.012 $\pm$ 0.002$\pm$ 0.003& 0.022 $\pm$ 0.002 $\pm$ 0.003\\
\bf{0.55} & 0.28 & 0.47 & 1.93 &0.67& 0.029 $\pm$ 0.002 $\pm$ 0.003 & -0.007 $\pm$ 0.003 $\pm$ 0.001& 0.012 $\pm$ 0.004 $\pm$ 0.006\\
\bf{0.65} & 0.29 & 0.48 & 1.93 &0.65& 0.039 $\pm$ 0.005 $\pm$ 0.000 & -0.017 $\pm$ 0.005 $\pm$ 0.002& 0.019 $\pm$ 0.005 $\pm$ 0.007\\
\bf{0.75} & 0.29 & 0.48 & 1.95 &0.65& 0.049 $\pm$ 0.005 $\pm$ 0.002 & -0.028 $\pm$ 0.007 $\pm$ 0.003& 0.013  $\pm$ 0.011 $\pm$ 0.005\\
\bf{0.85} & 0.30 & 0.44 & 1.95 &0.65& 0.052 $\pm$ 0.003 $\pm$ 0.003 & -0.041 $\pm$ 0.011$\pm$ 0.002& \\
\bf{0.95} & 0.30 & 0.29 & 1.93 &0.63& 0.066 $\pm$ 0.003 $\pm$ 0.004 & -0.049 $\pm$ 0.034 $\pm$ 0.008& \\
\hline
0.51 & \bf{0.14} & 0.52 & 1.27 &0.72& 0.018 $\pm$ 0.003 $\pm$ 0.004& -0.014 $\pm$ 0.006 $\pm$ 0.004& 0.012 $\pm$ 0.004 $\pm$ 0.011\\
0.51 & \bf{0.24} & 0.45 & 1.69 &0.65& 0.023 $\pm$ 0.003 $\pm$ 0.001& -0.008 $\pm$ 0.003 $\pm$ 0.002 & 0.016 $\pm$ 0.003 $\pm$ 0.003\\
0.51 & \bf{0.34} & 0.40 & 2.17 &0.65& 0.026 $\pm$ 0.003 $\pm$ 0.001& -0.008 $\pm$ 0.003 $\pm$ 0.003 & 0.023 $\pm$ 0.003 $\pm$ 0.006\\
0.51 & \bf{0.44} & 0.38 & 2.91 &0.65& 0.027 $\pm$ 0.002 $\pm$ 0.003& -0.010 $\pm$ 0.006 $\pm$ 0.003& 0.026 $\pm$ 0.005 $\pm$ 0.003\\
0.51 & \bf{0.54} & 0.36 & 3.78 &0.70& 0.026 $\pm$ 0.006 $\pm$ 0.002& -0.019 $\pm$ 0.015 $\pm$ 0.004& -0.015 $\pm$ 0.015 $\pm$ 0.007\\
\hline
0.52 & 0.32 & \bf{0.10} & 1.94 &0.65& 0.017 $\pm$ 0.004 $\pm$ 0.003 & 0.012 $\pm$ 0.008 $\pm$ 0.004 & 0.041 $\pm$ 0.010 $\pm$ 0.011\\
0.51 & 0.29 & \bf{0.30} & 1.95 &0.68& 0.028 $\pm$ 0.002 $\pm$ 0.003 & 0.013 $\pm$ 0.006 $\pm$ 0.002 & 0.031 $\pm$ 0.005 $\pm$ 0.005 \\
0.51 & 0.28 & \bf{0.50} & 1.91 &0.69& 0.024 $\pm$ 0.003 $\pm$ 0.003 & -0.012 $\pm$ 0.006 $\pm$ 0.003 & 0.026 $\pm$ 0.003 $\pm$ 0.002\\
0.51 & 0.26 & \bf{0.70} & 1.85 &0.70& 0.023 $\pm$ 0.004 $\pm$ 0.003 & -0.016 $\pm$ 0.005 $\pm$ 0.002 & 0.003 $\pm$ 0.004 $\pm$ 0.001\\
0.52 & 0.21 & \bf{0.90} & 1.69 &0.73& 0.029 $\pm$ 0.005 $\pm$ 0.003 & 0.010 $\pm$ 0.007 $\pm$ 0.005 & 0.016 $\pm$ 0.006 $\pm$ 0.002\\
\hline
0.51 & 0.21 & 0.46 & \bf{1.35} &0.68& 0.018 $\pm$ 0.002 $\pm$ 0.003 & -0.011 $\pm$ 0.002 $\pm$ 0.003& \\
0.51 & 0.30 & 0.43 & \bf{2.05} &0.67& 0.026 $\pm$ 0.004 $\pm$ 0.001 & -0.014 $\pm$ 0.003 $\pm$ 0.001& 0.011 $\pm$ 0.004 $\pm$ 0.007\\
0.51 & 0.37 & 0.42 & \bf{2.75} &0.71& 0.030 $\pm$ 0.002 $\pm$ 0.001 & -0.011 $\pm$ 0.005 $\pm$ 0.003& 0.021 $\pm$ 0.003 $\pm$ 0.001\\
0.51 & 0.45 & 0.40 & \bf{3.45} &0.73& 0.031 $\pm$ 0.004 $\pm$ 0.003 & -0.016 $\pm$ 0.009 $\pm$ 0.002& 0.023 $\pm$ 0.004 $\pm$ 0.005\\
0.51 & 0.52 & 0.37 & \bf{4.15} &0.75& 0.019 $\pm$ 0.005 $\pm$ 0.004 & -0.006 $\pm$ 0.019 $\pm$ 0.005& 0.012 $\pm$ 0.007 $\pm$ 0.006\\
\hline
\end{tabular}
\caption{$A_{LU}^{\sin\phi}$ in one dimension vs $z$, $x$, $P_T$ and $Q^2$. The dependent variable for each set is displayed in bold text.}
\label{tab:alu1d}
\end{table*}

\begin{table*}
\centering
\begin{tabular}{|c|c|c|c|c|c|c|}
\hline
 $<x>$ & $<P_{T}>$ & $<z>$ & $<Q^{2}>$ & $A_{LU}^{\sin\phi}$, $\pi^{+}$ $\pm (stat) \pm (syst)$& $A_{LU}^{\sin\phi}$, $\pi^{-}$ $\pm (stat) \pm (syst)$& $A_{LU}^{\sin\phi}$, $\pi^{0}$ $\pm (stat) \pm (syst)$\\
\hline
\bf{0.19} & \bf{0.19} & 0.54 & 1.20 & 0.030 $\pm$ 0.009 $\pm$ 0.009 &  &  \\
\bf{0.17 }& \bf{0.34} & 0.50 & 1.25 & 0.024 $\pm$ 0.004 $\pm$ 0.005 &  & 0.049 $\pm$ 0.012 $\pm$ 0.014 \\
\bf{0.16 }& \bf{0.51 }& 0.49 & 1.26 & 0.018 $\pm$ 0.004 $\pm$ 0.007 & -0.020 $\pm$ 0.009 $\pm$ 0.005 & 0.042 $\pm$ 0.011 $\pm$ 0.008 \\
\bf{0.16 }& \bf{0.69 }& 0.49 & 1.27 & 0.025 $\pm$ 0.004 $\pm$ 0.004 & -0.020 $\pm$ 0.008 $\pm$ 0.005 & 0.019 $\pm$ 0.017 $\pm$ 0.011 \\
\bf{0.16 }& \bf{0.88 }& 0.50 & 1.27 & 0.015 $\pm$ 0.005 $\pm$ 0.006 & 0.014 $\pm$ 0.011 $\pm$ 0.010 &  \\
\bf{0.26 }& \bf{0.16 }& 0.54 & 1.49 & 0.025 $\pm$ 0.004 $\pm$ 0.003 &  & 0.000 $\pm$ 0.000 $\pm$ 0.008 \\
\bf{0.25 }& \bf{0.32 }& 0.51 & 1.58 & 0.031 $\pm$ 0.002 $\pm$ 0.006 & 0.049 $\pm$ 0.011 $\pm$ 0.003 & 0.044 $\pm$ 0.007 $\pm$ 0.007 \\
\bf{0.24 }& \bf{0.50 }& 0.50 & 1.63 & 0.029 $\pm$ 0.002 $\pm$ 0.005 & -0.022 $\pm$ 0.005 $\pm$ 0.005 & 0.049 $\pm$ 0.007 $\pm$ 0.005 \\
\bf{0.24 }& \bf{0.69 }& 0.50 & 1.66 & 0.026 $\pm$ 0.003 $\pm$ 0.004 & -0.017 $\pm$ 0.005 $\pm$ 0.005 & 0.014 $\pm$ 0.010 $\pm$ 0.011 \\
\bf{0.23 }& \bf{0.87 }& 0.50 & 1.75 & 0.033 $\pm$ 0.004 $\pm$ 0.009 & -0.023 $\pm$ 0.010 $\pm$ 0.013 & 0.000 $\pm$ 0.000 $\pm$ 0.001 \\
\bf{0.34 }& \bf{0.14 }& 0.54 & 1.87 & 0.012 $\pm$ 0.003 $\pm$ 0.003 & 0.019 $\pm$ 0.014 $\pm$ 0.008 & 0.037 $\pm$ 0.013 $\pm$ 0.008 \\
\bf{0.33 }& \bf{0.31 }& 0.51 & 2.00 & 0.031 $\pm$ 0.004 $\pm$ 0.004 & 0.011 $\pm$ 0.006 $\pm$ 0.004 & 0.026 $\pm$ 0.006 $\pm$ 0.004 \\
\bf{0.33 }& \bf{0.50 }& 0.50 & 2.05 & 0.032 $\pm$ 0.003 $\pm$ 0.004 & -0.022 $\pm$ 0.006 $\pm$ 0.003 & 0.017 $\pm$ 0.004 $\pm$ 0.004 \\
\bf{0.33 }& \bf{0.67 }& 0.49 & 2.16 & 0.027 $\pm$ 0.003 $\pm$ 0.007 & -0.036 $\pm$ 0.007 $\pm$ 0.011 & 0.035 $\pm$ 0.010 $\pm$ 0.006 \\
\bf{0.32 }& \bf{0.85 }& 0.49 & 2.51 & 0.030 $\pm$ 0.006 $\pm$ 0.006 & -0.002 $\pm$ 0.013 $\pm$ 0.018 &  \\
\bf{0.43 }& \bf{0.14 }& 0.53 & 2.61 & 0.020 $\pm$ 0.005 $\pm$ 0.004 & -0.018 $\pm$ 0.027 $\pm$ 0.007 & -0.008 $\pm$ 0.008 $\pm$ 0.007 \\
\bf{0.43 }& \bf{0.31 }& 0.51 & 2.77 & 0.035 $\pm$ 0.004 $\pm$ 0.009 & -0.000 $\pm$ 0.012 $\pm$ 0.010 & -0.004 $\pm$ 0.004 $\pm$ 0.010 \\
\bf{0.43 }& \bf{0.50 }& 0.50 & 2.80 & 0.037 $\pm$ 0.006 $\pm$ 0.008 & -0.005 $\pm$ 0.008 $\pm$ 0.006 & 0.014 $\pm$ 0.008 $\pm$ 0.008 \\
\bf{0.43 }& \bf{0.66 }& 0.48 & 2.95 & 0.039 $\pm$ 0.006 $\pm$ 0.012 & -0.044 $\pm$ 0.010 $\pm$ 0.019 & 0.033 $\pm$ 0.013 $\pm$ 0.011 \\
\bf{0.41 }& \bf{0.83 }& 0.47 & 3.31 & 0.028 $\pm$ 0.010 $\pm$ 0.021 & -0.016 $\pm$ 0.025 $\pm$ 0.018 &  \\
\bf{0.52 }& \bf{0.15 }& 0.53 & 3.59 & 0.040 $\pm$ 0.013 $\pm$ 0.008 &  & 0.007 $\pm$ 0.011 $\pm$ 0.009 \\
\bf{0.52 }& \bf{0.31 }& 0.51 & 3.70 & 0.039 $\pm$ 0.010 $\pm$ 0.013 & -0.020 $\pm$ 0.029 $\pm$ 0.033 & 0.015 $\pm$ 0.011 $\pm$ 0.009 \\
\bf{0.52 }& \bf{0.50 }& 0.49 & 3.73 & 0.026 $\pm$ 0.008 $\pm$ 0.008 & -0.009 $\pm$ 0.021 $\pm$ 0.028 & 0.031 $\pm$ 0.026 $\pm$ 0.010 \\
\bf{0.52 }& \bf{0.64 }& 0.47 & 3.78 & 0.042 $\pm$ 0.013 $\pm$ 0.043 & -0.054 $\pm$ 0.026 $\pm$ 0.011 &   \\
\hline
\end{tabular}
\caption{$A_{LU}^{\sin\phi}$ binned two-dimensionally in  $x_B$ and $P_{T}$.}
\label{tab:alu2d}
\end{table*}


\section{Conclusion}

The $\sin\phi$ moment of the SIDIS cross section corresponding to a polarized lepton beam scattering from an unpolarized target, $A_{LU}^{\sin\phi}$, has been measured with absolute statistical accuracy of better than 1\% in all three pion channels. The measurements are compared to previous results published by the CLAS~\cite{e16, Avakian:2003pk, Aghasyan:2011ha} and HERMES~\cite{Airapetian:2006rx} Collaborations. This data represents an improvement of 2.6 times the HERMES precision and extends the kinematic coverage. This is most notable for $\pi^{-}$, where this is the first CLAS result and the previous result from HERMES does not have sufficient statistics to establish the sign of the asymmetry. The data reinforces the previous $\pi^{0}$ result from the CLAS Collaboration. That result was measured with a different beam energy and a modified detector configuration, so the current $\pi^{0}$ result is important primarily to reduce the risk of a systematic effect when comparing it with the other pion channels. The current measurements are in good agreement with all previous measurements, and serve as the first evidence for the negative sign of the $\pi^{-}$ $\sin\phi$ moment.

The 12 GeV upgrade at Jefferson Lab will provide the opportunity to further improve these measurements with higher statistics and a wider kinematic range, though it will be several years before this data is available~\cite{Avakian:2011zz}. Furthermore, the solid understanding of twist-3 fragmentation functions and TMDs will be very important for development of the physics case for future facilities such as the proposed Electron Ion Collider~\cite{eic}, and the current measurement will contribute to our understanding of these factors.

\begin{acknowledgements}
The authors of this paper would like to thank the staff of the Thomas Jefferson National Accelerator Facility who made this experiment possible. We owe much gratitude to P. Schweitzer for many fruitful discussions concerning the interpretation of our results. We also thank A. Prokudin, B. Pasquini, M. Schlegel, and Z. Lu for their insights. This work was supported in part by U.S Department of Energy grant DE-FG02-04ER41309. We are also thankful for the support of the U.S. National Science Foundation, the Italian Istituto Nazionale di Fisica Nucleare, the French Centre National de la Recherche Scientifique, the French Commissariat \`a l'\'Energie Atomique, the United Kingdom's Science and Technology Facilities Council,  the Chilean Comisi\'on Nacional de Investigaci\'on Cient\'ifica y Tecnol\'ogica, the Scottish Universities Physics Alliance, and the National Research Foundation of Korea. The Southeastern Universities Research Association (SURA) operated the Thomas Jefferson National Accelerator Facility for the US Department of Energy under Contract No. DE-AC05-84ER40150.
\end{acknowledgements}

\clearpage

\bibliographystyle{h-physrev}
\bibliography{sinphi_draft}

\begin{thebibliography}{10}

\bibitem{emc}
European Muon Collaboration, J.~Ashman {\em et~al.},
\newblock Nucl. Phys. {\bf B328}, 1 (1989).

\bibitem{collaboration:2011fga}
STAR Collaboration, P.~Djawotho,
\newblock (2011), arXiv:1106.5769.

\bibitem{Adamczyk:2012qj}
STAR Collaboration, L.~Adamczyk {\em et~al.},
\newblock Phys.Rev. {\bf D86}, 032006 (2012), 1205.2735.

\bibitem{Adare:2008aa}
PHENIX Collaboration, A.~Adare {\em et~al.},
\newblock Phys.Rev.Lett. {\bf 103}, 012003 (2009), 0810.0694.

\bibitem{deFlorian:2009vb}
D.~de~Florian, R.~Sassot, M.~Stratmann, and W.~Vogelsang,
\newblock Phys.Rev. {\bf D80}, 034030 (2009), 0904.3821.

\bibitem{Belitsky:2003nz}
A.~V. Belitsky, X.-d. Ji, and F.~Yuan,
\newblock Phys. Rev. {\bf D69}, 074014 (2004), hep-ph/0307383.

\bibitem{Meissner:2008xs}
S.~Meissner, A.~Metz, and M.~Schlegel,
\newblock p.~99 (2008), 0807.1154.

\bibitem{Lorce:2011dv}
C.~Lorce, B.~Pasquini, and M.~Vanderhaeghen,
\newblock JHEP {\bf 05}, 041 (2011), 1102.4704.

\bibitem{Bacchetta:2004jz}
A.~Bacchetta, U.~D'Alesio, M.~Diehl, and C.~A. Miller,
\newblock Phys. Rev. {\bf D70}, 117504 (2004), hep-ph/0410050.

\bibitem{Bacchetta:2006tn}
A.~Bacchetta {\em et~al.},
\newblock JHEP {\bf 02}, 093 (2007), hep-ph/0611265.

\bibitem{Levelt:1994np}
J.~Levelt and P.~J. Mulders,
\newblock Phys. Lett. {\bf B338}, 357 (1994), hep-ph/9408257.

\bibitem{Jaffe:1996zw}
R.~L. Jaffe,
\newblock (1996), hep-ph/9602236.

\bibitem{Efremov:2002ut}
A.~V. Efremov, K.~Goeke, and P.~Schweitzer,
\newblock Phys. Rev. {\bf D67}, 114014 (2003), hep-ph/0208124.

\bibitem{Schweitzer:2003uy}
P.~Schweitzer,
\newblock Phys. Rev. {\bf D67}, 114010 (2003), hep-ph/0303011.

\bibitem{Ohnishi:2003mf}
Y.~Ohnishi and M.~Wakamatsu,
\newblock Phys. Rev. {\bf D69}, 114002 (2004), hep-ph/0312044.

\bibitem{Afanasev:2003ze}
A.~Afanasev and C.~E. Carlson,
\newblock (2003), hep-ph/0308163.

\bibitem{Yuan:2003gu}
F.~Yuan,
\newblock Phys. Lett. {\bf B589}, 28 (2004), hep-ph/0310279.

\bibitem{Bacchetta:2004zf}
A.~Bacchetta, P.~J. Mulders, and F.~Pijlman,
\newblock Phys. Lett. {\bf B595}, 309 (2004), hep-ph/0405154.

\bibitem{Metz:2004je}
A.~Metz and M.~Schlegel,
\newblock Eur. Phys. J. {\bf A22}, 489 (2004), hep-ph/0403182.

\bibitem{Pijlman:2006vm}
F.~Pijlman,
\newblock (2006), hep-ph/0604226.

\bibitem{Lu:2012gu}
Z.~Lu and I.~Schmidt,
\newblock Phys. Lett. {\bf B712}, 451 (2012), 1202.0700.

\bibitem{Mao:2012dk}
W.~Mao and Z.~Lu,
\newblock Phys.Rev. {\bf D87}, 014012 (2013), 1210.4790.

\bibitem{Boer:1997nt}
D.~Boer and P.~J. Mulders,
\newblock Phys. Rev. {\bf D57}, 5780 (1998), hep-ph/9711485.

\bibitem{Sivers:1990fh}
D.~W. Sivers,
\newblock Phys.Rev. {\bf D43}, 261 (1991).

\bibitem{Jaffe:1991ra}
R.~Jaffe and X.-D. Ji,
\newblock Nucl. Phys. {\bf B375}, 527 (1992).

\bibitem{Burkardt:2009rf}
M.~Burkardt,
\newblock AIP Conf. Proc. {\bf 1155}, 26 (2009), 0905.4079.

\bibitem{Collins:1992kk}
J.~C. Collins,
\newblock Nucl. Phys. {\bf B396}, 161 (1993), hep-ph/9208213.

\bibitem{Seidl:2008xc}
Belle Collaboration, R.~Seidl {\em et~al.},
\newblock Phys. Rev. {\bf D78}, 032011 (2008), 0805.2975.

\bibitem{Airapetian:2010ds}
HERMES Collaboration, A.~Airapetian {\em et~al.},
\newblock Phys. Lett. {\bf B693}, 11 (2010), 1006.4221.

\bibitem{Alekseev:2010rw}
COMPASS Collaboration, M.~Alekseev {\em et~al.},
\newblock Phys. Lett. {\bf B692}, 240 (2010), 1005.5609.

\bibitem{Penttinen:2000dg}
M.~Penttinen, M.~V. Polyakov, A.~Shuvaev, and M.~Strikman,
\newblock Phys. Lett. {\bf B491}, 96 (2000), hep-ph/0006321.

\bibitem{Ji:2012ba}
X.~Ji, X.~Xiong, and F.~Yuan,
\newblock (2012), arXiv:1207.5221.

\bibitem{Hatta:2012cs}
Y.~Hatta and S.~Yoshida,
\newblock JHEP {\bf 1210}, 080 (2012), 1207.5332.

\bibitem{clasnim}
B.~A.~Mecking {\em et~al.},
\newblock Nucl. Instrum. Meth. {\bf A503}, 513 (2003).

\bibitem{dcnim}
M.~Mestayer {\em et~al.},
\newblock Nucl. Instrum. Meth. {\bf A449}, 81 (2000).

\bibitem{ccnim}
G.~Adams {\em et~al.},
\newblock Nucl. Instrum. Meth. {\bf A465}, 414 (2001).

\bibitem{ecnim}
M.~Amarian {\em et~al.},
\newblock Nucl. Instrum. Meth. {\bf A460}, 239 (2001).

\bibitem{tofnim}
E.~Smith {\em et~al.},
\newblock Nucl.Instrum.Meth. {\bf A432}, 265 (1999).

\bibitem{Schweitzer:2010tt}
P.~Schweitzer, T.~Teckentrup, and A.~Metz,
\newblock Phys. Rev. {\bf D81}, 094019 (2010), 1003.2190.

\bibitem{ROOT}
R.~Brun and F.~Rademaker,
\newblock Nucl. Instrum. Meth. {\bf A389}, 81 (1997).

\bibitem{Ingelman:1996mq}
G.~Ingelman, A.~Edin, and J.~Rathsman,
\newblock Comput. Phys. Commun. {\bf 101}, 108 (1997), hep-ph/9605286.

\bibitem{Brun:1978fy}
R.~Brun, R.~Hagelberg, M.~Hansroul, and J.~Lassalle,
\newblock (1978).

\bibitem{clasdb}
{CLAS physics database} ,
\newblock {\url{http://clasweb.jlab.org/physicsdb}}.

\bibitem{Airapetian:2006rx}
HERMES Collaboration, A.~Airapetian {\em et~al.},
\newblock Phys. Lett. {\bf B648}, 164 (2007), hep-ex/0612059.

\bibitem{e16}
CLAS Collaboration, H.~Avakian {\em et~al.},
\newblock High Energy Spin Physics Proc. , 239 (2003).

\bibitem{Avakian:2003pk}
CLAS Collaboration, H.~Avakian {\em et~al.},
\newblock Phys. Rev. {\bf D69}, 112004 (2004), hep-ex/0301005.

\bibitem{Aghasyan:2011ha}
M.~Aghasyan {\em et~al.},
\newblock Phys. Lett. {\bf B704}, 397 (2011), 1106.2293.

\bibitem{Efremov:2002sd}
A.~V. Efremov, K.~Goeke, and P.~Schweitzer,
\newblock Nucl. Phys. {\bf A711}, 84 (2002), hep-ph/0206267.

\bibitem{Avakian:2009jt}
CLAS Collaboration, H.~Avakian {\em et~al.},
\newblock Mod. Phys. Lett. {\bf A24}, 2995 (2009), 0910.3181.

\bibitem{Avakian:2010br}
H.~Avakian, A.~V. Efremov, P.~Schweitzer, and F.~Yuan,
\newblock (2010), 1001.5467.

\bibitem{Efremov:2006qm}
A.~V. Efremov, K.~Goeke, and P.~Schweitzer,
\newblock Phys. Rev. {\bf D73}, 094025 (2006), hep-ph/0603054.

\bibitem{Pasquini:2011tk}
B.~Pasquini and P.~Schweitzer,
\newblock Phys. Rev. {\bf D83}, 114044 (2011), 1103.5977.

\bibitem{Mao:2013waa}
W.~Mao and Z.~Lu,
\newblock Eur.Phys.J. {\bf C73}, 2557 (2013), arXiv:1306.1004.

\bibitem{Bacchetta:2008af}
A.~Bacchetta, F.~Conti, and M.~Radici,
\newblock Phys.Rev. {\bf D78}, 074010 (2008), 0807.0323.

\bibitem{Bacchetta:2003rz}
A.~Bacchetta, A.~Schaefer, and J.-J. Yang,
\newblock Phys.Lett. {\bf B578}, 109 (2004), hep-ph/0309246.

\bibitem{Airapetian:2009ti}
HERMES Collaboration, A.~Airapetian {\em et~al.},
\newblock Phys. Rev. Lett. {\bf 103}, 152002 (2009), 0906.3918.

\bibitem{Adolph:2012sp}
COMPASS Collaboration, C.~Adolph {\em et~al.},
\newblock Phys.Lett. {\bf B717}, 383 (2012), 1205.5122.

\bibitem{Adolph:2012sn}
COMPASS Collaboration, C.~Adolph {\em et~al.},
\newblock Phys.Lett. {\bf B717}, 376 (2012), 1205.5121.

\bibitem{Avakian:2011zz}
H.~Avakian,
\newblock AIP Conf. Proc. {\bf 1388}, 464 (2011).

\bibitem{eic}
D.~Boer {\em et~al.},
\newblock (2011), 1108.1713,
\newblock 547 pages, A report on the joint BNL/INT/Jlab program on the science
  case for an Electron-Ion Collider, September 13 to November 19, 2010,
  Institute for Nuclear Theory, Seattle.

\end{thebibliography}

\end{document}